\documentclass[a4paper,prd,showpacs,preprintnumbers,superscriptaddress,nofootinbib,preprint,floatfix]{revtex4-1}

\usepackage[utf8]{inputenc}
\usepackage{amsmath,amsfonts,amssymb,physics}
\usepackage{graphicx, hyperref}
\usepackage[colorinlistoftodos]{todonotes}

\begin{document}

\preprint{IPMU19-0119}

\title{Fragileness of Exact I-ball/Oscillon}

\author{Masahiro Ibe}
\email[e-mail: ]{ibe@icrr.u-tokyo.ac.jp}
\affiliation{ICRR, The University of Tokyo, Kashiwa, Chiba 277-8582, Japan}
\affiliation{Kavli IPMU (WPI), UTIAS, The University of Tokyo, Kashiwa, Chiba 277-8583, Japan}
\author{Masahiro Kawasaki}
\email[e-mail: ]{kawasaki@icrr.u-tokyo.ac.jp}
\affiliation{ICRR, The University of Tokyo, Kashiwa, Chiba 277-8582, Japan}
\affiliation{Kavli IPMU (WPI), UTIAS, The University of Tokyo, Kashiwa, Chiba 277-8583, Japan}
\author{Wakutaka Nakano}
\email[e-mail: ]{m156077@icrr.u-tokyo.ac.jp}
\affiliation{ICRR, The University of Tokyo, Kashiwa, Chiba 277-8582, Japan}
\affiliation{Kavli IPMU (WPI), UTIAS, The University of Tokyo, Kashiwa, Chiba 277-8583, Japan}
\author{Eisuke Sonomoto}
\email[e-mail: ]{sonomoto@icrr.u-tokyo.ac.jp}
\affiliation{ICRR, The University of Tokyo, Kashiwa, Chiba 277-8582, Japan}
\affiliation{Kavli IPMU (WPI), UTIAS, The University of Tokyo, Kashiwa, Chiba 277-8583, Japan}

\date{August 2019}

\begin{abstract}
I-ball/oscillon is a soliton-like oscillating configuration of a real scalar field which lasts for a long time.
I-ball/oscillon is a minimum energy state for a given adiabatic invariant, and its approximate conservation guarantees the longevity.
In this paper, we examine the stability of a special type of I-ball/oscillon, the ``exact" I-ball/oscillon, 
whose adiabatic invariant is exactly conserved.
We show that the exact I-ball/oscillon is stable in classical field theory,
but not stable against small perturbations depending on the value of its adiabatic invariant. 
Accordingly, the exact I-ball/oscillon breaks up 
in the presence of the fluctuations
with corresponding instability modes.
We also confirm the fragileness of the exact I-ball/oscillon by the classical lattice simulation.
\end{abstract}

\maketitle

\section{Introduction}

I-ball/oscillon is a non-topological soliton-like solution in real scalar field theory~\cite{Bogolyubsky:1976yu,Gleiser:1993pt,Copeland:1995fq} whose formation process is nonlinear.
This lump of a real scalar field is a minimum energy state and understood as the coherent oscillation around the (local) minimum of the potential.
A lot of studies on the I-ball/oscillon 
\cite{McDonald:2001iv,Amin:2010jq,Amin:2011hj,Amin:2013ika,Takeda:2014qma,Lozanov:2016hid,Hasegawa:2017iay,Antusch:2017flz,Hong:2017ooe,Kawasaki:2015vga,Fodor:2006zs,Fodor:2008du,Gleiser:2008ty,Fodor:2009kf,Gleiser:2009ys,Hertzberg:2010yz,Salmi:2012ta,Saffin:2014yka,Kawasaki:2013awa,Mukaida:2016hwd,Eby:2018ufi,Ibe:2019vyo,Olle:2019kbo,Antusch:2019qrr}
indicate that the lifetime of I-ball/oscillon is extremely long.
The longevity of the I-ball/oscillon could leave an 
imprint on cosmology/astrophysics which can be tested 
by some experiments. (See e.g.~\cite{Zhou:2013tsa,Antusch:2016con,Lozanov:2019ylm}
for gravitational waves from the I-ball/oscillon. See also \cite{Eby:2019ntd}.)

The longevity of I-ball/oscillon is guaranteed by the approximate conservation of adiabatic invariant $I$~\cite{Kasuya:2002zs,Kawasaki:2015vga}. 
The adiabatic invariance is the analog to the invariance of the phase space volume for a periodic motion in classical mechanics.
It is also known that the adiabatic invariant corresponds to the particle number~\cite{Mukaida:2014oza} 
in the non-relativistic limit, and hence, the approximate conservation of $I$ 
is regarded as the particle number conservation in this limit.

In this paper, we study the stability of a special type of the I-ball/oscillon, the ``exact" I-ball/oscillon, 
which appears in a real scalar theory with a particular type of potential~\cite{Kawasaki:2015vga}.
In this particular case, the adiabatic invariant of I-ball/oscillon is exactly conserved.
We show that the exact I-ball/oscillon is stable in classical field theory.
We also find that the exact I-ball/oscillon is not stable against small perturbations  depending on the value of the adiabatic invariant. 
Accordingly, the exact I-ball/oscillon breaks up 
in the presence of the fluctuations
corresponding to the instability modes.
We also confirm the fragileness of the exact I-ball/oscillon by the classical lattice simulation.

Organization of this paper is as follows.
In section~\ref{sec:I-ball}, we introduce the exact I-ball/oscillon, which conserves the adiabatic invariant exactly.
In section~\ref{sec:perturb}, we show that the exact I-ball/oscillon is stable,
but the perturbation around it has resonance bands and it breaks the exact I-ball/oscillon depending on its parameters.
In section~\ref{sec:sim}, we show the setup and the result of our lattice simulation to confirm fragileness of the exact I-ball/oscillon.
Finally, in section~\ref{sec:conclusion}, we conclude the results.

\section{Exact I-ball/oscillon}
\label{sec:I-ball}


\subsection{Exact conservation of adiabatic invariant}

I-ball/oscillon is a localized oscillating scalar field configuration which minimizes the energy for a given value of the adiabatic invariant $I$.
The adiabatic invariant of a real scalar field $\phi$ 
is defined by
\begin{equation}
	I = \frac{1}{\omega}\int d^3x \overline{\dot{\phi}^2}\ ,
    \label{eq:adiabatic_inv}
\end{equation}
where $\omega$ is the angular frequency of the oscillating field
and the overbar denotes the average over one period of the oscillation.

The adiabatic invariant Eq.~(\ref{eq:adiabatic_inv}) is approximately conserved
when the scalar field oscillates in the potential dominated by the quadratic term ($\sim\phi^2$).
In other words, the I-ball/oscillon is not an exact periodic motion in time,
and hence, the time average over one period of the oscillation in Eq.\,\eqref{eq:adiabatic_inv} is not exact.
Due to this approximate conservation of $I$,
I-ball/oscillon is generally quasi-stable and eventually decays by emitting scalar waves~\cite{Ibe:2019vyo}.

However, the adiabatic invariant is exactly conserved 
when the solution of the equation of motion is completely separable into the time and the spatial dependent parts~\cite{Kawasaki:2015vga}.

Let us assume an I-ball/oscillon solution with the separated form as
\begin{equation}
   \phi(t,\vec{x}) = f(t)\psi(\mathbf{x}).
   \label{eq:separable}
\end{equation}
where $\psi(\mathbf{x})$ is the theoretical oscillon profile and 
$f(t)$ is a time periodic function normalized as $\max\{f(t)\} = 1$.
The fact that the time and spatial dependencies are determined separately is crucial
when we consider the I-ball/oscillon stability.
In this case, the adiabatic invariant $I$ is evaluated as
\begin{align}
    I = \frac{\overline{\dot f^2}}{\omega}\int d^3x \,\psi(\mathbf x)^2\ ,
    \label{eq:I}
\end{align}
where overbar denotes the time average over the one period of oscillation.
Because $f(t)$ is exactly periodic,
the adiabatic invariant $I$ is constant in time, 
and hence, is conserved exactly.

Such a separated solution like Eq.~(\ref{eq:separable}) is possible
only when the scalar potential of $\phi$ takes the form of
\begin{equation}
    V= \frac{1}{2} m^2 \phi^2 + \Delta V
     = \frac{1}{2} m^2 \phi^2
         +\frac{1}{2}\kappa m^2 \phi^2 \log\frac{\phi^2}{M^2}\ .
    \label{eq:potential}
\end{equation}
where $\kappa<0$ is a dimensionless constant and $m$ and $M$ are mass parameters
\footnote{
The parameter $M$ corresponds to the renormalization scale.
}
\cite{Kawasaki:2015vga}.

For a later purpose, we redefine the parameters by
\begin{eqnarray}
\tilde m^2 &=& m^2 + \kappa m^2 \log\frac{\tilde m^2}{M^2} \ , \\
\tilde \kappa \tilde m^2 &=& \kappa m^2 \ ,
\end{eqnarray}
with which 
\begin{equation}
    V = \frac{1}{2} \tilde{m}^2 \phi^2
         +\frac{1}{2}\tilde \kappa \tilde{m}^2 \phi^2 \log\frac{\phi^2}{\tilde m^2}\ .
    \label{eq:potential2}
\end{equation}
In what follows, we use the latter expression with $\tilde m^2 \to m^2$ and $\tilde \kappa \to \kappa$.
Eventually, the potential depends only on two parameters, $m$ and $\kappa$.

Substituting the solution Eq.~(\ref{eq:separable}) into the equation of motion
\begin{equation}
    \ddot{\phi} -\nabla \phi + m^2 \phi + \Delta V' = 0\ ,
    \label{eq:eom}
\end{equation}
we obtain
\begin{equation}
    \left[\ddot{f} + m^2 f + \kappa m^2 \left(1+ \log f^2\right)f\right]\psi
    -\left[\nabla \psi - \kappa m^2 \psi \log\frac{\psi^2}{m^2}\right]f = 0\ .
\end{equation}
This leads to the following two equations:
\begin{align}
    \ddot{f} + m^2 f + \kappa m^2 \left(1+ \log f^2\right) f
    & =  \zeta m^2 f \ ,
    \label{eq:eom_f}\\
    \nabla^2 \psi - \left(\kappa m^2 \log\frac{\psi^2}{m^2}\right)\psi
    & =  \zeta m^2 \psi\ ,
    \label{eq:eom_psi}
\end{align}
where $\zeta$ is a constant.
As we will see in the next section, $\zeta$ determines 
the adiabatic invariant of the I-ball/oscillon for 
given potential parameters.

Therefore, because the solutions of equations of motions under the potential Eq.\,(\ref{eq:potential}) are independently determined by Eqs.~(\ref{eq:eom_f}) and (\ref{eq:eom_psi})
the adiabatic invariant is exactly conserved
\footnote{
From Eq.~(\ref{eq:eom_f}),  we can evaluate $\overline{\dot{f}^2}$ as
\begin{align}
    \overline{\dot{f}^2}
     = m^2 \left(1+\kappa -\zeta\right)\overline{f^2}
       + \kappa m^2 \,\overline{f^2 \log f^2}
       \ .
       \label{eq:f_ave_rel}
\end{align}
}
as shown in Ref.~\cite{Kawasaki:2015vga}.

\subsection{The exact I-ball/oscillon solution}

As the I-ball/oscillon corresponds to the minimum energy states 
for a given value of $I$, the I-ball/oscillon profile 
$\psi$ is expected to be spherical,  $\psi(\vec{x})=\psi(r)$. Then, Eq.\,\eqref{eq:eom_psi} has an exact solution,
\begin{equation}
    \psi(r) = \psi_c \exp(-r^2/R^2)\ ,
    \label{eq:gaussian_sol}
\end{equation}
where 
\begin{align}
    R =& \sqrt{\frac{2}{|\kappa|}}\frac{1}{m}\ , \\
    \psi_c^2 =& m^2 \exp \left(3-\frac{\zeta}{\kappa}\right)\ .
    \label{eq:cond_psic}
\end{align}
$\psi_c$ and $R$ is the central value and the radius of the I-ball/oscillon.
It should be noted that the spatial size of $R$ does not depend on $\zeta$, and hence, does not depend on $I$
for given potential parameters.

\begin{figure}[t]
    \begin{minipage}{.4\linewidth}
        \begin{center}
            \includegraphics[width=\linewidth]{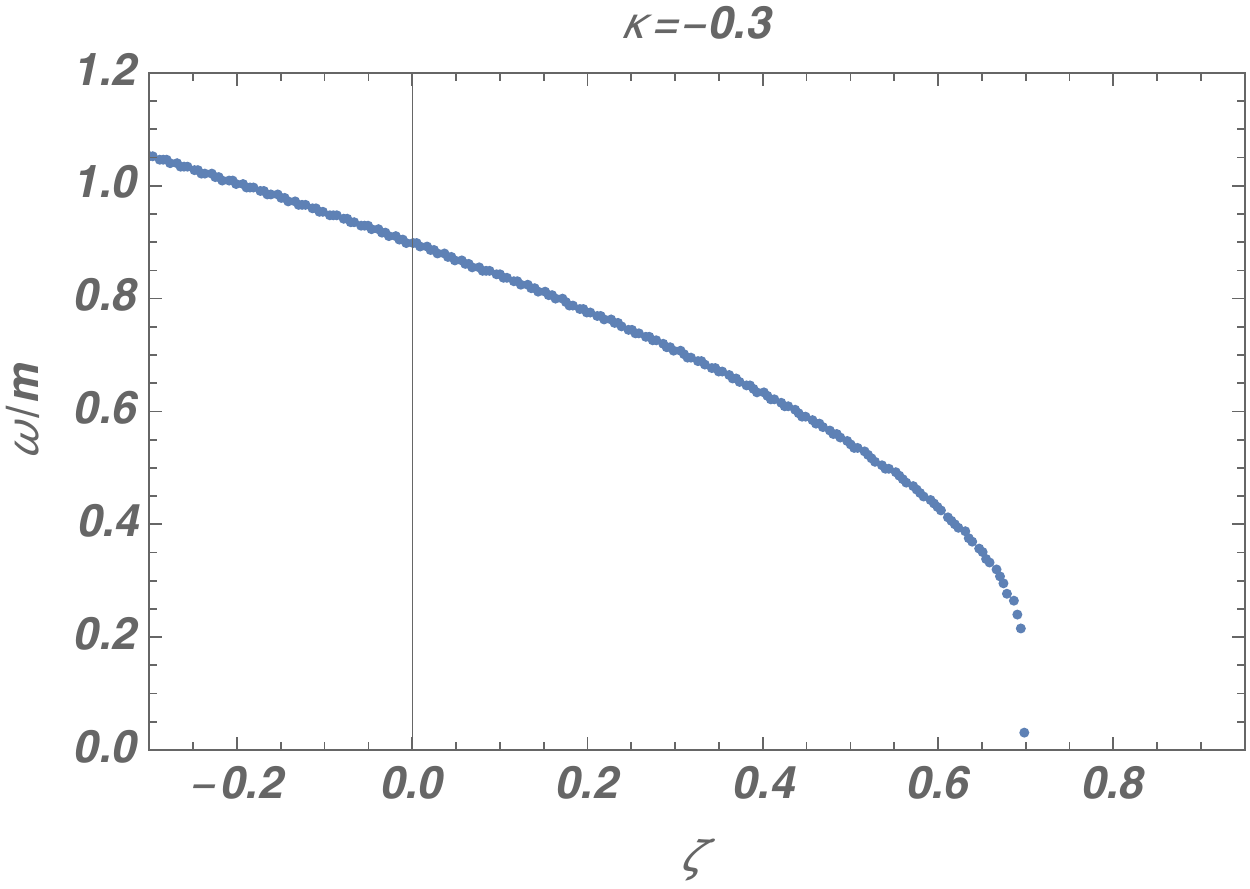}
        \end{center}
    \end{minipage}
    \hspace{.3cm}
    \begin{minipage}{.4\linewidth}
        \begin{center}
            \includegraphics[width=\linewidth]{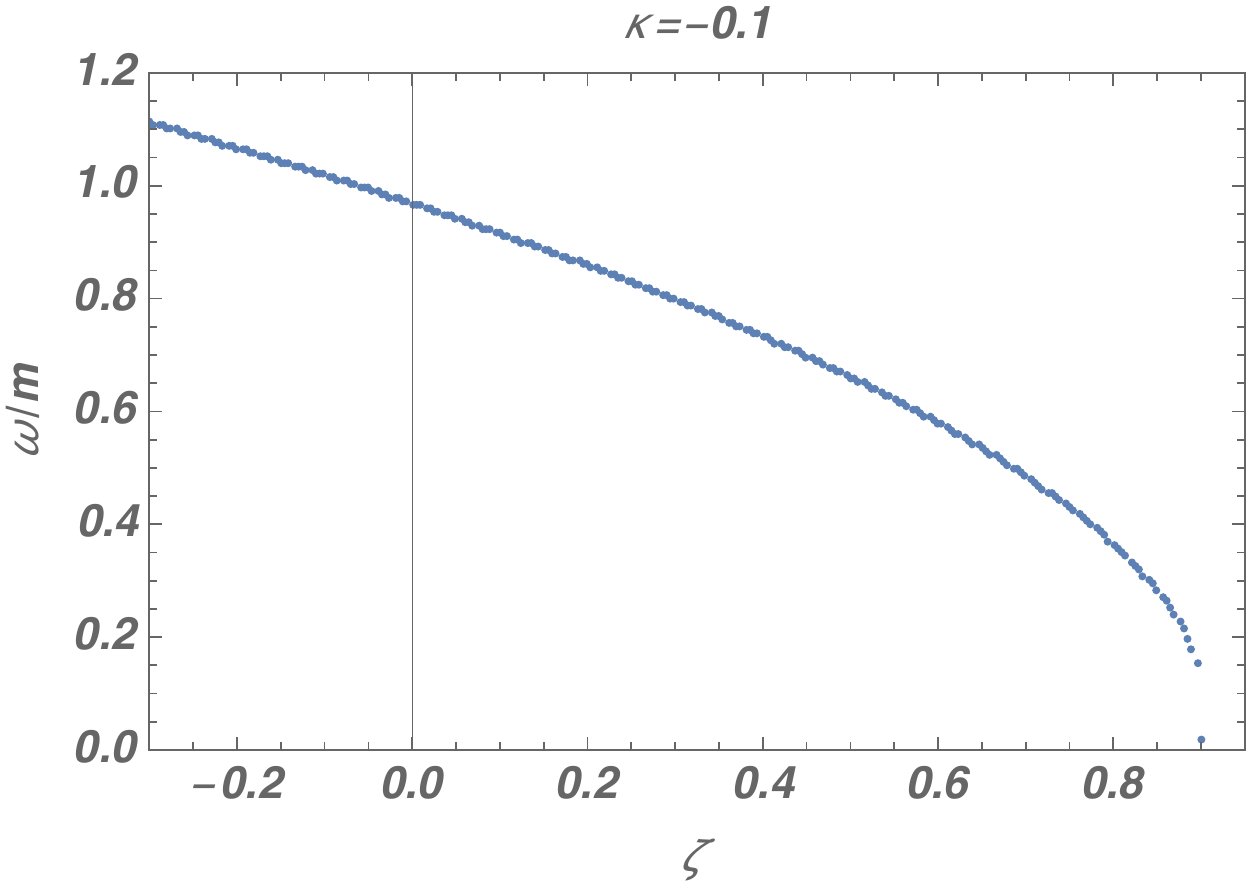}
        \end{center}
    \end{minipage}
    \caption{\sl (Left) The relation of $\zeta$ and $\omega$ for $\kappa=-0.3$.
    (Right) Same plot for $\kappa=-0.1$.
    }
    \label{fig:zeta_omega}
\end{figure}

The period of the oscillation can be obtained as follow.
By multiplying $\dot f$ to Eq.\,(\ref{eq:eom_f}) and integrating over time $t$, we obtain
\begin{eqnarray}
\frac{1}{2}\dot f^2
+ \frac{1}{2}m^2 (1-\zeta) f^2 + \frac{1}{2}\kappa m^2 f^2 \log f^2
= C_f\ ,
\end{eqnarray}
with $C_f$ being a constant.
Because the potential of $f$ is a even function of $f$,
$f$ oscillates in $f = [-1,1]$,
and hence, $C_f$ is required to be $C_f > 0$ to allow $\dot f^2 > 0$ at $f = 0$.
As we defined $f = \pm 1$ at the turning points of the motion,
$C_f$ is represented by
\begin{eqnarray}
C_f = \frac{1}{2} m^2 (1-\zeta) \ ,
\end{eqnarray}
with which $\dot f = 0$ at the turning points.
As a result, we find a period of the oscillation to be
\begin{align}
T &\equiv \frac{2\pi}{\omega} = 2\int_{-1}^1 \frac{ df}{\sqrt{ m^2(1-\zeta) (1-f^2) - \kappa m^2 f^2 \log f^2}}\ ,
\label{eq:period}
\end{align}
which is determined by $\zeta$ for given potential parameters.
We plot the frequencies for the potential parameter $\kappa=-0.3$ and $\kappa=-0.1$ in Fig.~\ref{fig:zeta_omega}.
The upper bound for the range of $\zeta$ is determined from Eq.\,\eqref{eq:period} with $T=$ finite,
\begin{align}
\zeta_{\mathrm{max}} = 1+\kappa \le 1 \ .
\end{align}
Besides, the region of $\omega \ge m$ is not physically attractive\footnote{More precise lower bound on $\zeta$ is determined by ${dE}/{dI} < m $.
}, and hence, we consider only closed range of $\zeta$.

Similarly, the energy of the I-ball/oscillon is given by
\begin{align}
E = & \left(\overline{\dot f^2}
-\frac{1}{2} \kappa m^2 \overline{f^2}   
 \right) \int d^3 x \,\psi(r)^2 \ , \\
 = & \left(1 -  \frac{1}{2} \kappa m^2 \frac{ \overline{f^2} }{ \overline{\dot f^2} } \right)  \omega I \ ,
\end{align}
where 
\begin{align}
    I = \frac{\overline{\dot f^2}}{\omega}\int d^3x \psi(r)^2
      = \frac{\overline{\dot f^2}}{m \omega} \left(\frac{\pi}{|\kappa|}\right)^{3/2}
        \frac{\psi_c^2}{m^2}\ .
\end{align}
Here, we used Eqs.\,\eqref{eq:eom_f}, \eqref{eq:I}, and \eqref{eq:f_ave_rel}.
Notice that $\overline{\dot{f}^2}$, $\omega$ and $\psi_c$ depend on $\zeta$ through Eqs.(\ref{eq:f_ave_rel}), (\ref{eq:cond_psic}) and (\ref{eq:period}).
Thus, the adiabatic invariant is determined only by $\zeta$ for given model parameters $\kappa$ and $m$.

\section{Stability and Instability of the Exact I-ball/Oscillon}
\label{sec:perturb}

\subsection{Stability}
As shown in \cite{Ibe:2019vyo}, the non-exact I-ball/oscillon 
decays by emitting relativistic radiation of the scalar field.
Here, let us discuss whether the exact I-ball/oscillon obeys 
this decay process.

To see whether a relativistic radiation is emitted from the exact I-ball/oscillon, we consider a small perturbation 
around the I-ball/oscillon, $\phi(x) = \phi_I(x) + \xi(x)$.
Here, $\phi_I(x)$ is the I-ball/oscillon solution obtained in
the previous section.
Then, the equation of motion of the perturbation $\xi(x)$ is 
given by
\begin{align}
    (\Box+V'' )|_{\phi = \phi_I}\,\xi(x) = \order{\xi(x)^2} \ ,
    \label{eq:fluct_eom}
\end{align}
where
\begin{align}
    V''=m^2 \left( 1 + 3\kappa +  \kappa \log\frac{\phi^2}{m^2}\right)\ .
\end{align}
Eq.\,\eqref{eq:fluct_eom} shows that 
the perturbation does not have source terms, and 
the I-ball/oscillon solution is stable if $\xi=0$ initially.
Thus, unlike the case of the non-exact I-ball/oscillon,  
the exact I-ball/oscillon does not decay by emitting 
relativistic radiations.

The absence of the source term of $\xi$ stems from the fact that the equation for the I-ball/oscillon profile is the same as the equation of motion Eqs.\,(\ref{eq:eom_f}) and (\ref{eq:eom_psi}) without averaging over one period of the oscillation.
This contrasts with generic I-ball/oscillon solutions.
In general cases, the I-ball/oscillon solutions satisfy the equation of motion only after averaging over a period, and hence, the perturbation  has source terms coming from the potential which disappear if averaged.
This leads to the decay of the I-ball/oscillon as shown in Ref.~\cite{Ibe:2019vyo}.
These observations are consistent with the exactness of the conservation of the adiabatic invariant.

\subsection{Fragileness}
In this subsection, we discuss the instability of the exact I-ball/oscillon against small perturbations and 
derive the growth index (the Floquet exponent) of the 
instability.

As $\kappa < 0$, the perturbation around the vacuum, i.e. $\phi = 0$, has an infinite mass, i.e. $m_\xi^2 = \infty > 0$.%
\footnote{
The mass at the origin of this potential is divergent because 
\begin{align*}
    V''(\phi) = m^2\left[ (1+3\kappa) + \kappa \log \left( \frac{\phi}{m} \right)^2 \right] \rightarrow \infty. \ \ \ (\phi\rightarrow \pm 0)
\end{align*}
}
Thus, the perturbation around the vacuum is never excited.
Around the I-ball/oscillon solution, 
\begin{align}
    \phi_I(x) = f(t)\psi(r)\ ,  \quad
    \quad
    \psi(r) = \psi_c e^{-r^2/R^2}\ ,
\end{align}
on the other hand, the perturbation has a finite non-derivative kernel, 
\begin{align}
    \ddot{\xi}(x)-\nabla^2 \xi(x) 
    + \left(2|\kappa|m^2 \frac{r^2}{R^2} +  (1+3 \kappa) m^2 + \kappa m^2 \log\frac{\psi_c^2}{m^2} + F(t)\right)\xi(x)=0\ ,
    \label{eq:eom_perturb}
\end{align}
where $F(t) = \kappa m^2 \log f(t)^2$.
and we neglected $\order{\xi^2}$ term in Eq.\,\eqref{eq:fluct_eom}.
Thus, there could be instability modes around the I-ball/oscillon solution.

To analyze Eq.~(\ref{eq:eom_perturb}), let us remember that the eigenequation,
\begin{align}
\label{eq:3DHO}
    \left(-\nabla^2 + \omega_\xi^2\,r^2 \right)\lambda_{\bf n}({\bf x}) = 2 E_{\bf n} \lambda_{\bf n}({\bf x})\ ,
\end{align}
has the eigenfunctions and eigenvalues,
\begin{align}
\label{eq:lambda}
    \lambda_{\bf n}({\bf x}) &= \prod_{i=1}^3 
    \left(\frac{1}{n_i!\, 2^{n_i}}\sqrt{\frac{\omega_\xi}{\pi}}\right)^{1/2} H_{n_i}(\sqrt{\omega_\xi}x_i) e^{-\omega_
    \xi x_i^2/2}\ , \\
    2E_{\bf n} &= 2\omega_\xi (n_1 + n_2 + n_3 + 3/2)\ .
\end{align}
where $H_n$ denotes the Hermite polynomial of order $n$,
and the eigenfunctions satisfy  $\lambda_{\mathbf n}(|\mathbf x|\to \infty)=0$.

By expanding $\xi(x)$ by,
\begin{align}
    \xi(t, {\bf x}) = \sum_{\bf n} q_{\bf n}(t)\lambda_{\bf n}({\bf x}) \ ,
\end{align}
we find that each $q_{\bf n}(t)$ satisfies 
\begin{align}
\label{eq:perturbation}
    \left[
\partial_t^2 + 2 \omega_{\xi}(n_1+n_2+n_3)
+ \Lambda + F(t)
    \right]q_{\bf n}(t) = 0 \ .
\end{align}
Here, we defined
\begin{align}
    \omega_\xi^2 &= 2 |\kappa|\frac{m^2}{R^2} = \kappa^2 m^4\ , \\
    \Lambda &= (1+3 \kappa) m^2 + \kappa m^2 \log\frac{\psi_c^2}{m^2} + 3 \omega_\xi\ , \\
    &= (1+3 \kappa - \zeta) m^2 \ ,
\end{align}
where we used Eq.\,\eqref{eq:cond_psic}.
The exponential factor in Eq.\,\eqref{eq:lambda} is identical to that of the I-ball/oscillon, i.e.
\begin{align}
    e^{-\omega_\xi (x_1^2 + x_2^2 + x_3^2)/2} = e^{-r^2/R^2}\ .
\end{align}
If Eq.\,\eqref{eq:perturbation} has growing modes with non-trivial Floquet exponents,
the I-ball/oscillon solution can be unstable.

It should be noted that the perturbation of the zero mode, ${\bf n} = 0$, is redundant.
As the radius of the I-ball/oscillon does not depend on $\zeta$ but is determined by $\kappa$ and $m$, 
the addition of the zero mode perturbation just enhances the value of $I$,
which leads to an I-ball/oscillon with a slightly larger $I$. 

Because we confine ourselves to the spherical configuration in our numerical simulation,
we consider the spherical perturbations to analyze the result of our simulation.

Let us rewrite Eq.\,\eqref{eq:3DHO} in the spherical coordinate,
\begin{align}
    \left(-\frac{d^2}{dr^2} - \frac{2}{r} \frac{d}{dr}  + \frac{\ell(\ell+1)}{r^2}+\omega_\xi r^2 \right)R_{n_r,\ell}(r) = 2 E_{n_r,\ell} R_{n_r,\ell}(r) \ ,
\end{align}
where the eigenfunction is given by $R_{n_r,\ell}(r)Y_m^{\ell}(\Omega)$ with $Y_m^{\ell}(\Omega)$
being the spherical harmonics.
By introducing $\tilde r = \sqrt{\omega_\xi}\, r$,
the eigenfunction of the radial direction and the corresponding energy are given by,
\begin{align}
    R_{n_r,\ell}(\tilde r) &= \frac{2^{1/2}\omega_\xi^{3/2}\Gamma(n_r + \ell + 3/2)^{1/2}}{\Gamma(n_r+1)^{1/2}\Gamma(\ell+3/2)}\,\tilde{r}^\ell e^{-\tilde r^2/2}\,{}_1F_1\left(-n_r;{3}/{2}+\ell; \tilde {r}^2\right)\ ,\\
    2E_{n_r,\ell} &= 2(2n_r + \ell + 3/2 )\ ,
\end{align}
for $(n_r =0,1,2\cdots)$ and $(\ell = 0,1,2\cdots)$.
Here, $_1F_1(a;b;z)$ denotes the confluent hypergeometric function,
and the wave function is normalized so that
\begin{align}
    \int_0^\infty dr r^2 R_{n_r,
    \ell}(\sqrt{\omega_\xi} r) R_{n_r',\ell}(\sqrt{\omega_\xi} r)= \delta_{n_r n_r'}\ .
\end{align}

By expanding the perturbation by   
\begin{align}
    \xi(t,{\bf x}) = \sum_{n_r,\ell,m_z} q_{n_r, \ell, m_z}(t) R_{n_r,\ell}(r) Y_{\ell}^{m_z}(\Omega)\ ,
\end{align}
the $q_{n_r, \ell, m_z}(t)$ satisfies,
\begin{align}
\label{eq:perturbation2}
    \left[
\partial_t^2 + 2 \omega_{\xi}(2 n_r + \ell)
+ \Lambda + F(t)
    \right]q_{n_r,\ell,m_z}(t) = 0 \ .
\end{align}
Since we are interested in the effects of the spherical perturbation,
we hereafter take $\ell =0$ and denote $q_{n_r}(t) = q_{n_r,0,0}(t)$.

\begin{figure}[t]
    \begin{minipage}{.4\linewidth}
        \begin{center}
            \includegraphics[width=\linewidth]{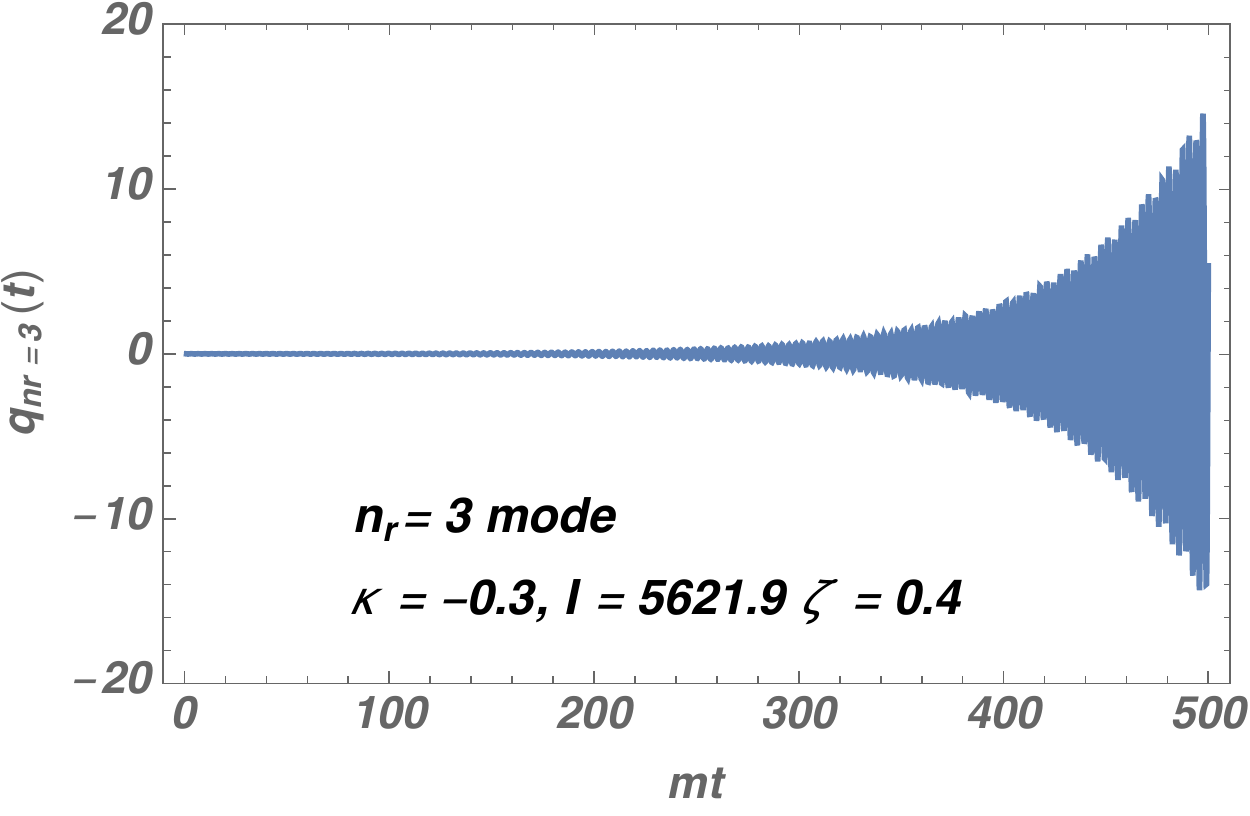}
        \end{center}
    \end{minipage}
    \hspace{.3cm}
    \begin{minipage}{.4\linewidth}
        \begin{center}
            \includegraphics[width=\linewidth]{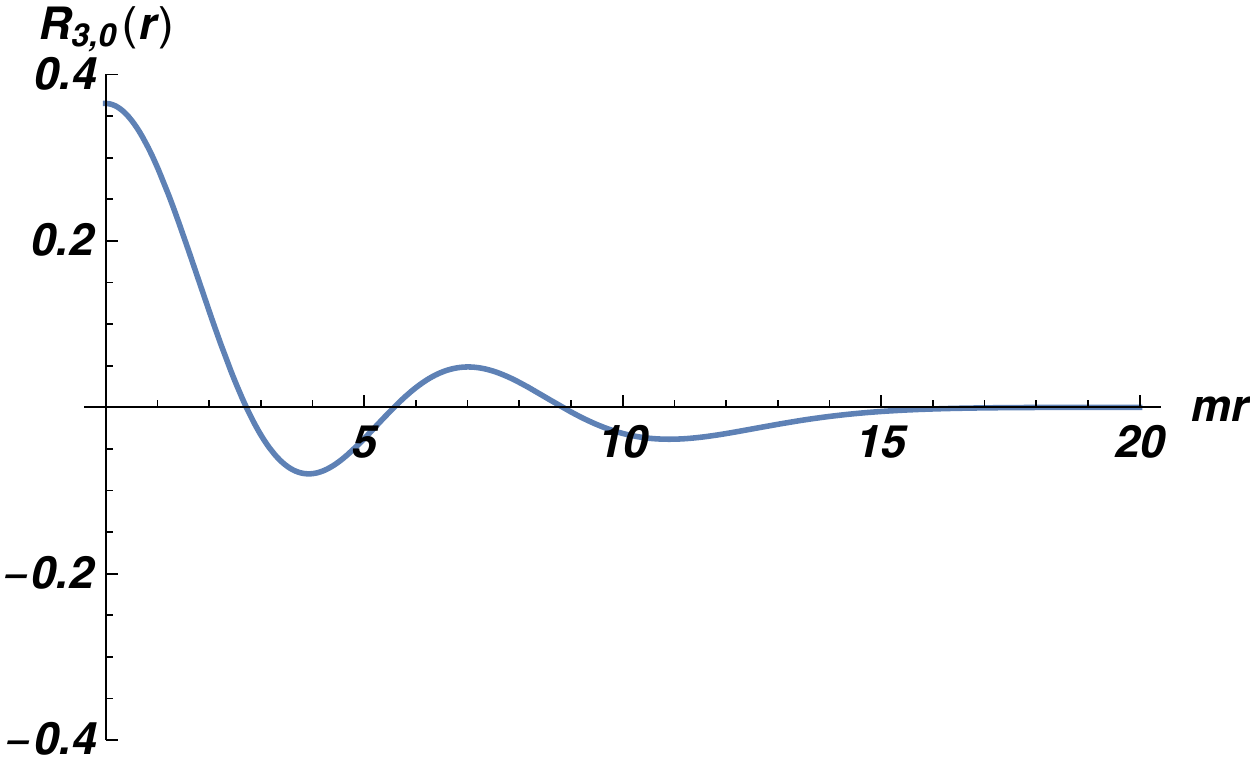}
        \end{center}
    \end{minipage}
    \caption{\sl (Left) The growing mode around the I-ball with $\kappa = -0.3$
    and $\zeta = 0.4$.
    (Right) The corresponding wave function $R_{n_r =3, 0}(r)$.
    }
    \label{fig:growth}
\end{figure}

In Fig.\,\ref{fig:growth}, we show the instability mode, $n_r = 3$,
for $\kappa = -0.3$ and $\zeta =0.4$.
The figure shows that the perturbation grows for ${\cal O}(100)/m$.
We also show the corresponding wave function $R_{n_r =3, 0}(r)$.
Once the perturbation grows around the I-ball/oscillon, it is expected to be dissociated and breaks up into 
smaller configurations.
 
\begin{figure}[th]
    \begin{minipage}{.6\linewidth}
        \begin{center}
            \includegraphics[width=\linewidth]{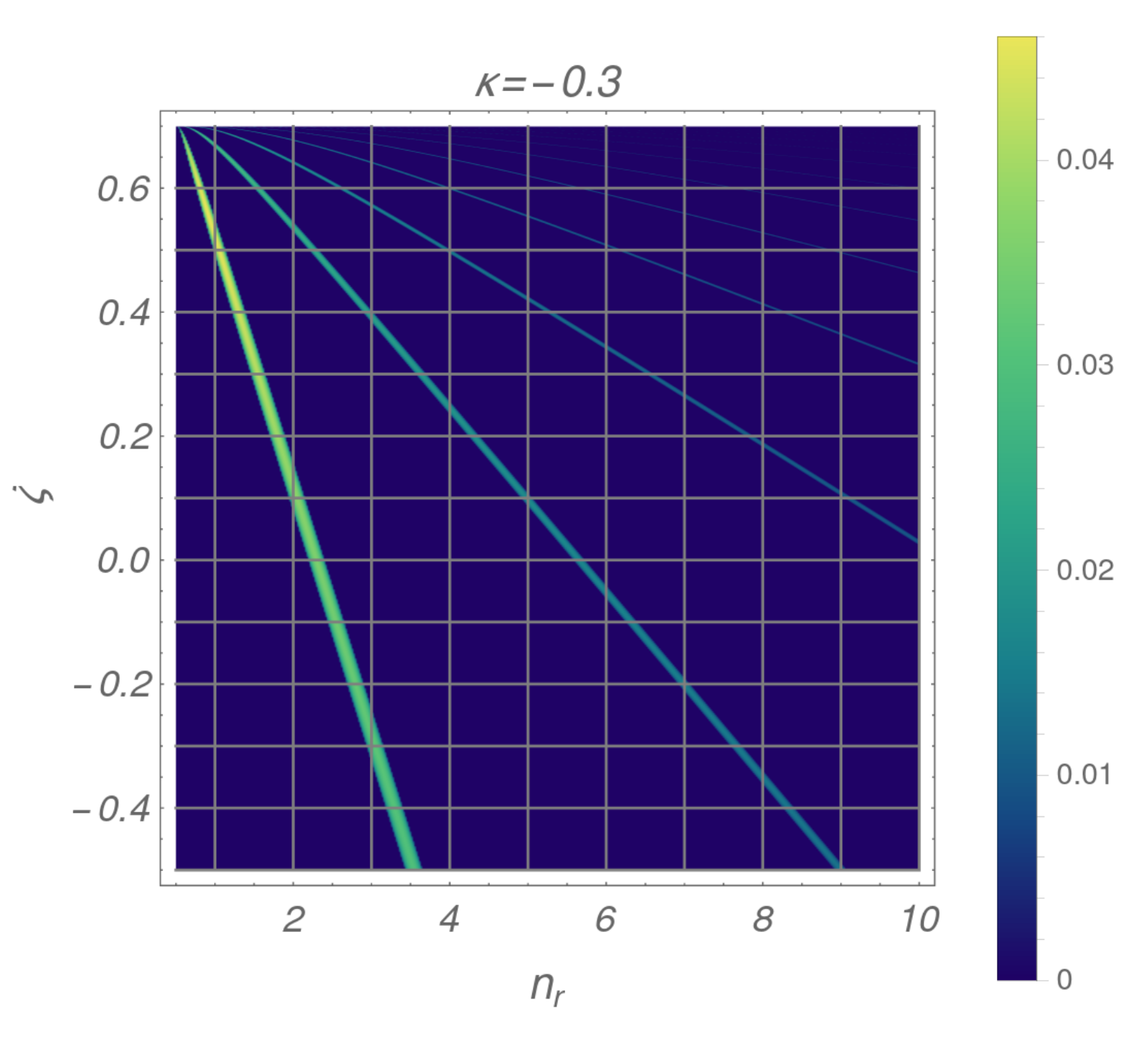}
        \end{center}
    \end{minipage}
    \hspace{.3cm}
    \begin{minipage}{.6\linewidth}
        \begin{center}
            \includegraphics[width=\linewidth]{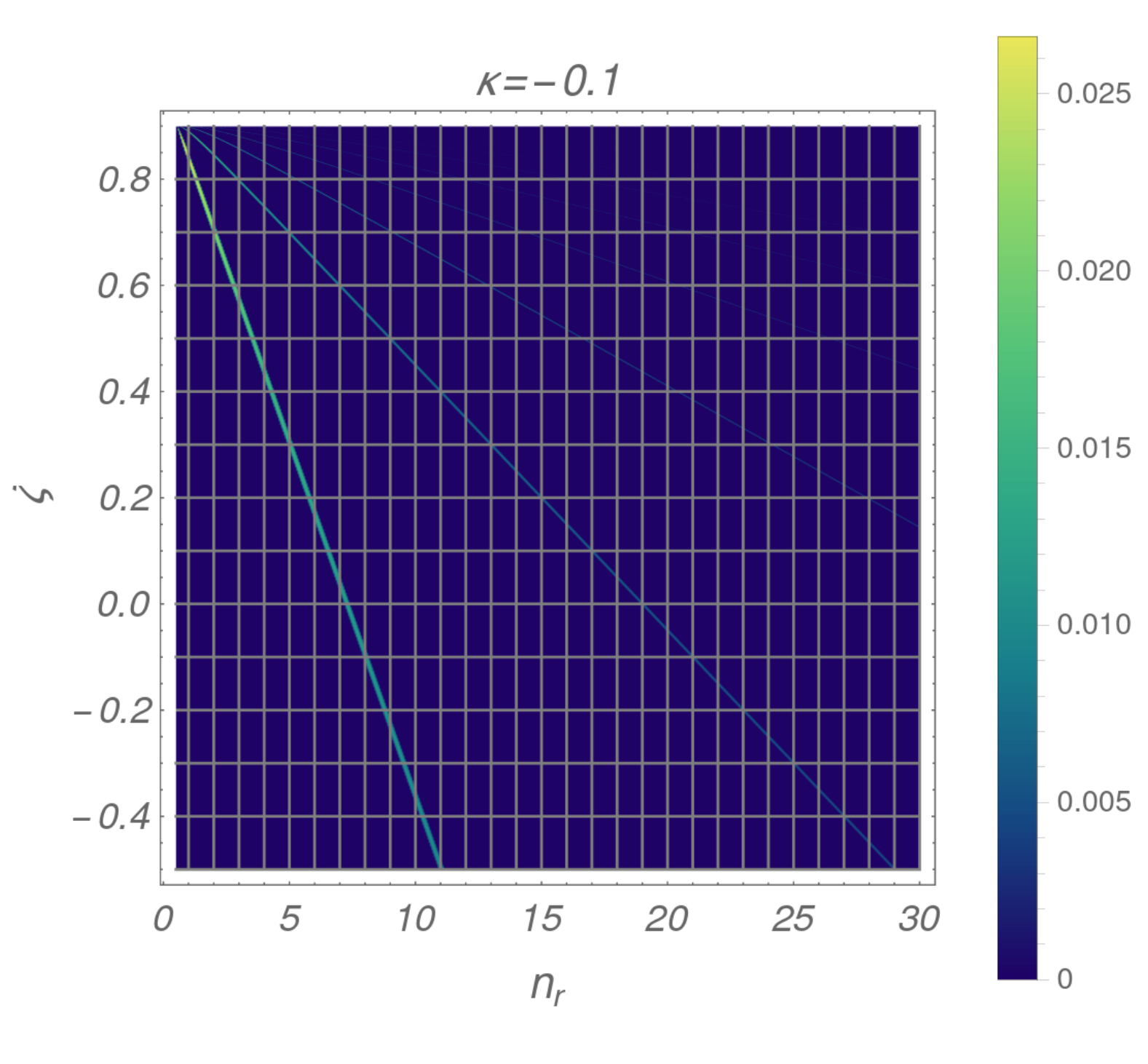}
        \end{center}
    \end{minipage}
    \caption{\sl
    (upper) Plot of the Floquet exponent $\mu$ for the variables $n_r$ and $\zeta$ for $\kappa = -0.3$.
    (lower) Same plot for $\kappa = -0.1$.
    The instability bands for the spherical modes are at integer values of $n_r$.
    The instability bands for the non-spherical modes
    can be read off by renaming the horizontal axis
    from $n_r$ to $n_r+\ell/2$.  }
    \label{fig:band}
\end{figure}

In Fig.\,\ref{fig:band}, we also show the Floquet exponent $\mu$
\begin{align}
    \xi(t+T,{\bf x}) = e^{\mu T} \xi(t,{\bf x})\ ,
\end{align}
where $T=2\pi/\omega$ is a period of one oscillation.
The figure shows that there are some resonance bands.
This means that the exact I-ball/oscillon is not always a stable solution.
Thus, in the presence of tiny fluctuations around the 
exact I-ball/oscillon, the instability modes exhibit 
the exponential growths in the time scale of $\order{\mu^{-1}}$.
Once the exponential growth happens, the exact I-ball/oscillon
cannot keep its configuration anymore and is expected to be 
broken up.
We will confirm these behaviors by the classical lattice simulation
in the next section.

\section{Numerical Simulation}
\label{sec:sim}


\subsection{Setup}
\label{sec:sim_setup}

In this subsection we briefly explain the setup of the simulation.
The procedure of the simulation is similar to that of Ref.~\cite{Ibe:2019vyo}. 
Because the lowest energy configuration of the scalar field $\phi$ is spherically symmetric in three-dimensional space,
the equation of motion of $\phi$ is represented by
\begin{equation}
    \frac{d^2 \phi}{d t^2}
    = \frac{d^2 \phi}{d r^2} + \frac{2}{r}\frac{d \phi}{d r} - \frac{\partial V}{\partial \phi}\ ,
    \label{eq:num_eom}
\end{equation}
where
\begin{align}
    V (\phi) = m^2 \left[ 1 + \kappa \log \left(\epsilon + \frac{\phi^2}{m^2} \right) \right] \phi^2\ .
\end{align}
Here, we introduced a small parameter $\epsilon$  to avoid the numerical instability 
caused by the singularity of the effective mass at $\phi = 0$.
We have confirmed that the simulation results are independent of this small regularization term $\epsilon$.

For the boundary condition, we use the two following conditions.
\begin{itemize}
\item At the origin $r=0$, to avoid the divergence of the second term of the right-hand side of Eq.\,(\ref{eq:num_eom}), we impose
\begin{equation}
    \left. \frac{1}{r}\frac{d \phi}{d r} \right|_{r=0} = 0\,.
\end{equation}
\item At the other boundary $r = L\ (\gg R)$, we impose the absorbing boundary condition (see Appendix \ref{sec:apdx_abc} for details).
Under this condition, the radiation of the real scalar field emitted from the I-ball/oscillon is absorbed at the boundary so that we correctly calculate the time evolution of I-ball/oscillon.
\end{itemize}

As the initial condition of $\phi$,
we use the theoretical I-ball/oscillon configuration Eqs.~(\ref{eq:gaussian_sol})-(\ref{eq:cond_psic}) 
for a given $\zeta_{\rm ini}$ with $1\%$ random fluctuations.
We also set 
\begin{eqnarray}
    \dot{\phi}(t=0,r) = 0\ , 
\end{eqnarray}
as an initial condition of $\dot{\phi}$.
We have confirmed that the exact I-ball/oscillon is completely stable in the absence 
of the random fluctuations.

\begin{table}[t]
    \centering
    \begin{tabular}{cc}
        \hline \hline
        $\zeta_{\rm ini}$ & varying \tabularnewline
        $\kappa$ & $-0.3$ \tabularnewline
        $\epsilon$ & $10^{-10}$ \tabularnewline
        Box size $L$  & $64$ \tabularnewline
        Grid size $N$ & $1024$ \tabularnewline
        Initial time  & $0$ \tabularnewline
        Final time    & $1.0\times 10^6$ \tabularnewline
        Time step     & $2.0\times 10^{-3}$ \tabularnewline
        \hline \hline
    \end{tabular}
    \caption{
        Simulation parameters.
        $\zeta_{\rm ini}$ is changed in every simulation to set the appropriate initial profile of the exact I-ball/oscillon.
    }
    \label{Ta:sim_para}
\end{table}
The other simulation parameters are shown in Table~\ref{Ta:sim_para}.
Here, the units of the field, time, space, etc. are taken to be $m^{-1}$, that is,
\begin{equation}
    \phi \rightarrow m\phi\ ,\ \
    t \rightarrow \frac{t}{m}\ ,\ \
    x \rightarrow \frac{x}{m}\ ,\ \dots\ {\rm etc}\ .
\end{equation}

We utilize the same lattice simulation code in \cite{Ibe:2019vyo},
in which the time evolution is calculated by the fourth-order symplectic integration scheme and the spatial derivatives are calculated by the fourth-order central difference scheme.

\subsection{Result}
\label{sec:sim_result}

We numerically calculate the time evolution of the exact I-ball/oscillon $\phi$ and its energy $E$ from
\begin{equation}
    E (t) =  \int ^L _0 dr\ 4\pi r^2
             \left[ \frac{1}{2}\dot{\phi}^2 + \frac{1}{2} (\nabla \phi)^2 + V \right],
    \label{eq:num_E}
\end{equation}
to see how it behaves in the presence of the instability modes derived in Sec.~\ref{sec:perturb}.

\subsubsection*{Stability}

\begin{figure}[t]
    \begin{center}
        \includegraphics[width=95mm]{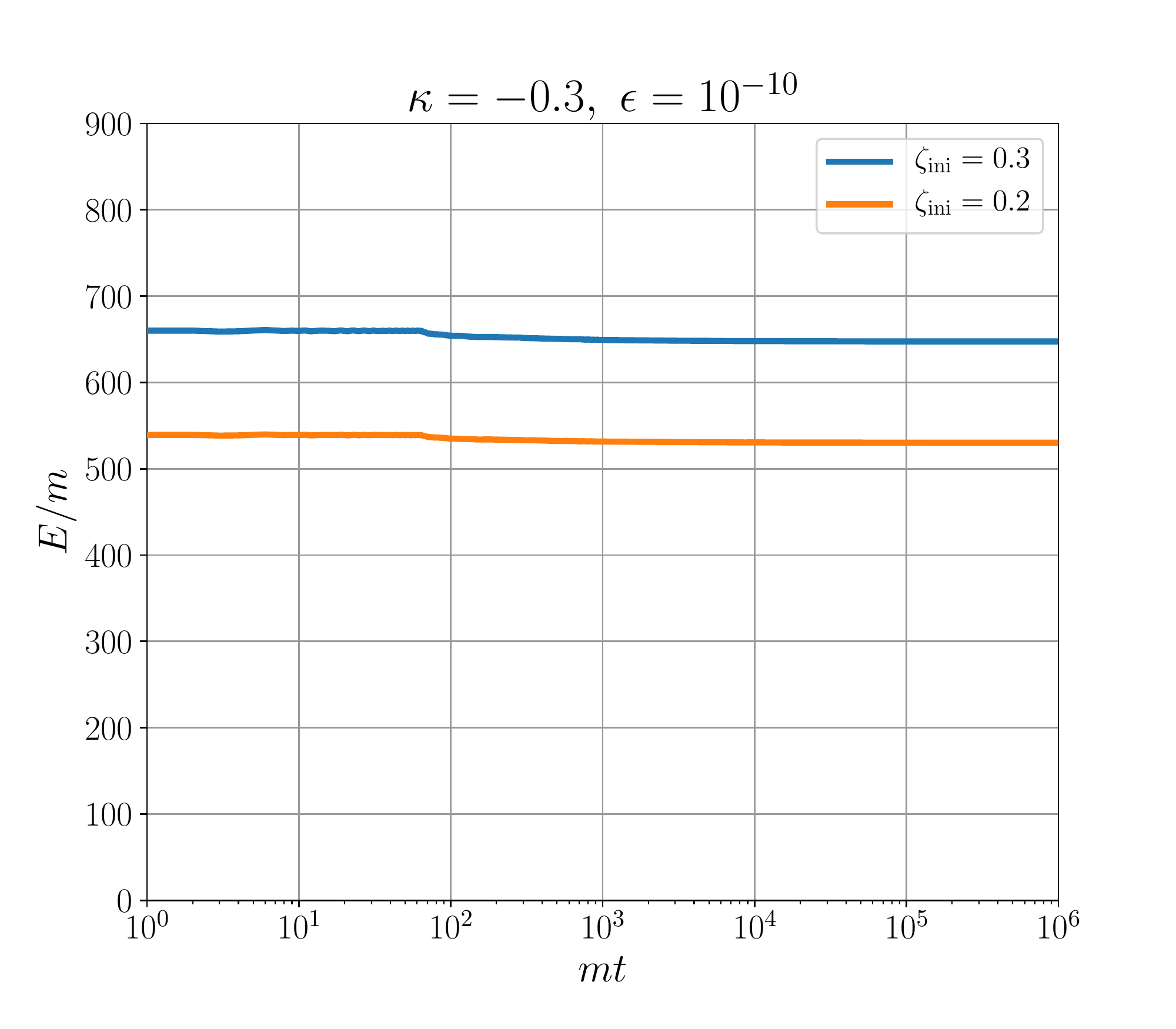}
    \end{center}
    \vspace{-1cm}
    \caption{\sl
        The result of our simulations for the stable exact I-ball/oscillon.
        The blue and the orange lines show the exact I-ball/oscillon energy (defined by Eq.\,(\ref{eq:num_E})) for $\zeta_{\rm ini} = 0.3$ and $\zeta_{\rm ini} = 0.2$, respectively.
        The energy suddenly changes at $mt \simeq 60$ 
        because the radiated fluctuations have reached the boundary and start to be absorbed at that time.
        We find that the exact I-ball/oscillon energy is conserved in $mt \lesssim 10^6$
        when the exact I-ball/oscillon does not have instability modes around it.
    }
    \label{fig:num_stable}
\end{figure}

First, we show the result of the stable exact I-ball/oscillon
which has no strong instability resonance bands.
The result is given in Fig.\,\ref{fig:num_stable}, which shows that the exact I-ball/oscillation 
does not break up  even in the presence of the tiny fluctuation.
This result is consistent with the fact the exact I-ball/oscillon for $\zeta_{\rm ini} = 0.3,\ 0.2$ 
does not have the instability modes.

\subsubsection*{Fragileness}

\begin{figure}[t]
    \begin{center}
        \includegraphics[width=125mm]{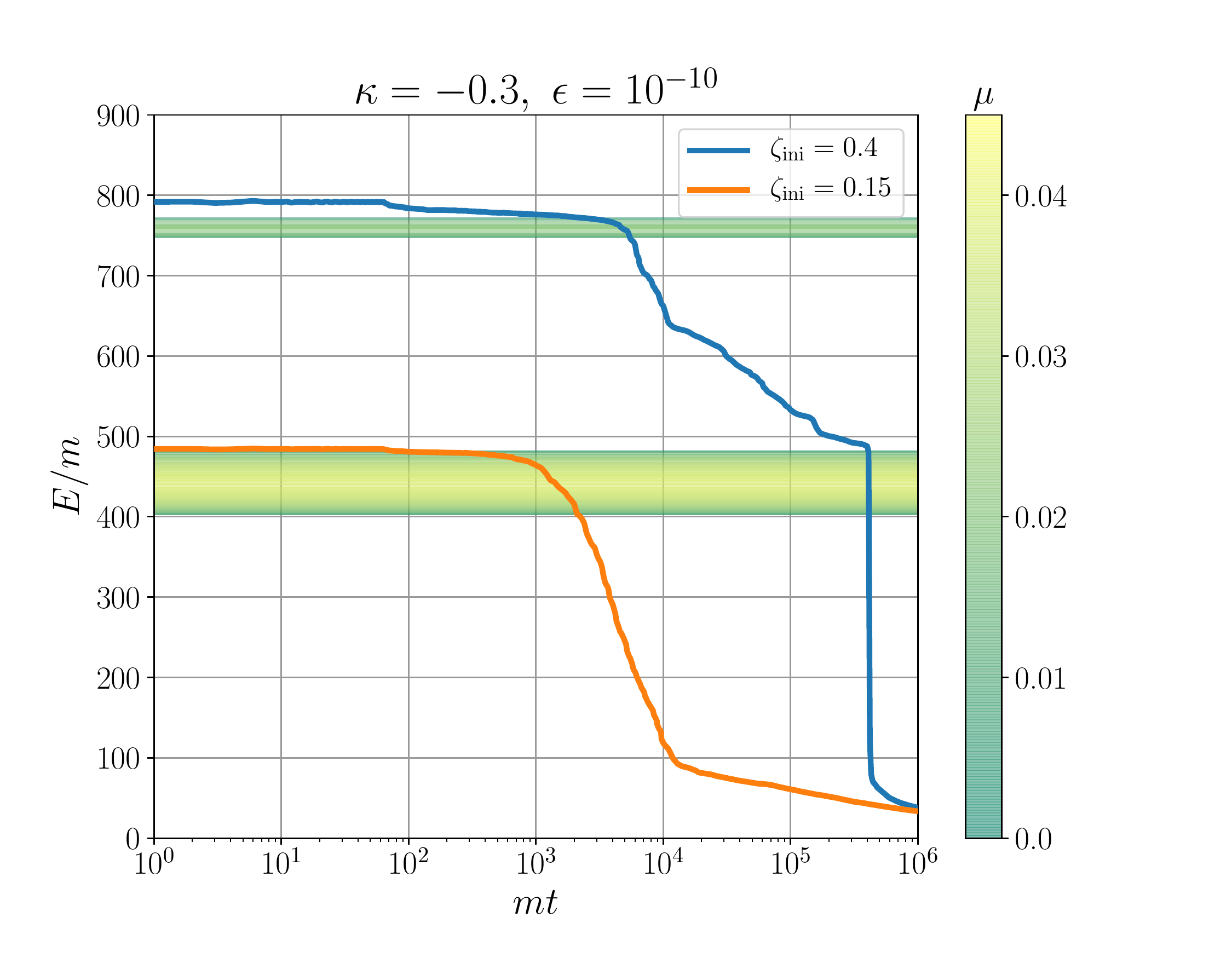}
    \end{center}
    \vspace{-1cm}
    \caption{\sl
        The result of our simulations for the unstable exact I-ball/oscillon, which hit the instability bands.
        The blue and the orange lines show the exact I-ball/oscillon energy (defined by Eq.\,(\ref{eq:num_E})) for $\zeta_{\rm ini} = 0.4$ and $\zeta_{\rm ini} = 0.15$ respectively.
        We also plot the typical instability bands as the Floquet exponent $\mu$ exhibited in Fig.~\ref{fig:band} by green thick lines.
        The upper dark green line and the lower bright green line correspond to the instability bands at $\zeta \simeq 0.4,\ n_r = 3$ and $\zeta \simeq 0.1,\ n_r = 2$ in Fig.~\ref{fig:band}.
        The sudden change of the energy at $mt \simeq 60$ is for the same reason as the case of Fig.~\ref{fig:num_stable}.
        We find that the exact I-ball/oscillon energy strikingly decreases around the instability bands.
    }
    \label{fig:num_unstable}
\end{figure}

Next, we show the result of the unstable exact I-ball/oscillon.
The results of the simulations are shown in Fig.\,\ref{fig:num_unstable}.%
\footnote{We have confirmed that the exact I-ball/oscillon for the given $\zeta_{\rm ini}$ without initial fluctuations is stable within simulation time.}
Comparing the result with our analytical calculation (see Fig.\,\ref{fig:band}),
we find that the energy of exact I-ball/oscillon strikingly decreases at around instability bands showed as green color bands in Fig.\,\ref{fig:num_unstable}.
This can be interpreted that the initial fluctuations appended to the exact I-ball/oscillon grow exponentially 
and deform the exact I-ball/oscillon profile
\footnote{
Instability bands should be wide and strong enough for the oscillon decay.
}.


Our results also suggest that the exact I-ball/oscillon
ends up with another exact I-ball/oscillon profile
with a smaller $I$ after the decay process.
In fact, both the cases in Fig.\,\ref{fig:num_unstable}
converge to the smaller but finite energy at $mt \gtrsim 10^6$.
The case with $\zeta_{\rm ini} = 0.4$ is particularly suggestive.
In this case, the I-ball/oscillon first decays 
when the instability modes in the narrower band around 
$\zeta = 0.4$ and $n_r = 3$ grows,
and ends up with the I-ball/oscillon with $\zeta \simeq 0.15$.
However, there is a broader instability band for $\zeta \simeq 0.15$ and $n_r = 2$, with which the the exact 
I-ball/oscillon further decays very quickly.
As a result, the exact I-ball/oscillon profile for 
$\zeta_{\rm ini} = 0.4$ converges to the smaller I-ball/oscillon solution.

In our analysis, we only consider the radial modes of the fluctuation.
The exact I-ball/oscillon can have more instability bands 
in the three dimensional case (see Eqs.\,\eqref{eq:perturbation} and \eqref{eq:perturbation2}).

\section{Conclusions}
\label{sec:conclusion}
In this paper, we examine the stability of the exact I-ball/oscillon.
The exact I-ball/oscillon has been considered to be stable since 
it has the exactly conserved adiabatic invariant.
Its stability is also expected because the perturbations around the exact I-ball/oscillon obey the field equation without the source terms.

However, we have found that the exact I-ball/oscillon is not 
always a stable solution and the perturbation around it has 
growth modes depending on the size of the adiabatic invariant.
Thus, in the presence of the fluctuations with the corresponding instability modes,
the exact I-ball/oscillon cannot keep its configuration anymore and breaks up eventually.

The mechanism of the exact I-ball/oscillon decay in this paper is completely different from the previous study~\cite{Ibe:2019vyo},
in which I-ball/oscillon decays by emitting relativistic radiations of the scalar field.
Our results in this paper suggest that it is necessary to consider both decay processes,
the decay by radiation and the decay by the instability modes, 
when we estimate the lifetime of generic I-ball/oscillon.

\begin{acknowledgments}

This work was supported by JSPS KAKENHI Grant Nos. 17H01131 (M.K.) and 17K05434 (M.K.),
MEXT KAKENHI Grant Nos. 15H05889 (M.K., M.I), No. 16H03991(M.I.), No. 17H02878(M.I.), and No. 18H05542 (M.I.),
World Premier International Research Center Initiative (WPI Initiative), MEXT, Japan,
and JSPS Research Fellowships for Young Scientists Grant No. 19J12936 (E.S.).

\end{acknowledgments}

\appendix


\section{Absorbing Boundary Condition}
\label{sec:apdx_abc}

In this appendix, we briefly explain the Absorbing Boundary Condition (ABC) based on the references~\cite{d4443ca6ddba410f8a155c4dd67936d0, Salmi:2012ta}.
We only consider the single real scalar field $\phi$ for simplicity.

The equation of motion of $\phi$ is written as
\begin{equation}
\ddot{\phi}-\Delta\phi+\frac{\partial V}{\partial\phi}=0
\end{equation}
where $V$ is an arbitrary potential of $\phi$. 
Assuming that the field $\phi$ can be expanded around the minimum $\phi = v$ as
\begin{equation}
\phi=v+\varphi,\ \ \ (\varphi \ll v)
\end{equation}
where
\begin{equation}
\left.\frac{\partial V}{\partial\phi}\right|_{\phi=v}=0,
\end{equation}
the equation of motion of $\phi$ is rewritten as
\begin{equation}
\ddot{\varphi}-\Delta\varphi+m_{{\rm eff}}^{2}\varphi \simeq 0.
\end{equation}
where
\begin{align}
\left.\frac{\partial^{2}V}{\partial\phi^{2}}\right|_{\phi=v}\equiv m_{{\rm eff}}^{2}.
\end{align}
We ignore the higher terms of $\varphi$ because we consider $\varphi$ is much smaller than $v$
\footnote{
    The potential of $\phi$ is expanded as
    \begin{align}
        V\left(\phi\right) & =V\left(v+\varphi\right),\\
        & =V\left(v\right)+\left.\frac{\partial V}{\partial\phi}\right|_{\phi=v}\varphi+\frac{1}{2}\left.\frac{\partial^{2}V}{\partial\phi^{2}}\right|_{\phi=v}\varphi^{2}+\mathcal{O}\left(\varphi^{3}\right),\\
        & =V\left(v\right)+\frac{1}{2}\left.\frac{\partial^{2}V}{\partial\phi^{2}}\right|_{\phi=v}\varphi^{2}+\mathcal{O}\left(\varphi^{3}\right).
    \end{align}
}.

Assuming the spherical symmetry in three dimensions $\varphi= \varphi\left(r\right)$, the equations of motion becomes
\begin{equation}
\ddot{\varphi}-\frac{\partial^{2}\varphi}{\partial r^{2}}-\frac{2}{r}\frac{\partial\varphi}{\partial r}+m_{{\rm eff}}^{2}\varphi \simeq 0.
\end{equation}
Performing Fourier-transformation of $\varphi$ as $t$,
\begin{equation}
\varphi\left(t,r\right)=\int d\omega\varphi\left(\omega,r\right)e^{-i\omega t}.
\end{equation}
the equations of motion becomes
\begin{align}
\left[\frac{\partial^{2}}{\partial r^{2}}+\frac{2}{r}\frac{\partial}{\partial r}+\left(\omega^{2}-m_{{\rm eff}}^{2}\right)\right]\varphi & =0,\\
\Leftrightarrow\ \ \ \left[\frac{\partial}{\partial r}-\left(i\sqrt{\omega^{2}-m_{{\rm eff}}^{2}}-\frac{1}{r}\right)\right]\left[\frac{\partial}{\partial r}+\left(i\sqrt{\omega^{2}-m_{{\rm eff}}^{2}}-\frac{1}{r}\right)\right]\varphi & =0.
\end{align}
The special solution of this equation at $m_{{\rm eff}}r \gg 1$ is Bessel function and
\begin{equation}
\varphi\left(t,r\right)=\frac{1}{r}\exp\left[i\left(\pm\sqrt{\omega^{2}-m_{{\rm eff}}^{2}}r+\omega t\right)\right].
\end{equation}
The reflected wave must be disappeared at the boundary $r = r_{b}$ $(m_{{\rm eff}}r \gg r_{b})$, 
\begin{equation}
\left.\left[\frac{\partial}{\partial r}-\left(i\sqrt{\omega^{2}-m_{{\rm eff}}^{2}}-\frac{1}{r}\right)\right]\right|_{r=r_{b}}\varphi=0.
\end{equation}
Expanding the square root,
\begin{align}
\left[\frac{\partial}{\partial r}-i\omega\left(1-\frac{1}{2}\frac{m_{{\rm eff}}^{2}}{\omega^{2}}+\mathcal{O}\left(\left(\frac{\omega}{m_{{\rm eff}}}\right)^{4}\right)\right)+\frac{1}{r}\right]\varphi & =0,\\
\Leftrightarrow \ \ \left[i\omega\frac{\partial}{\partial r}+\left(\omega^{2}-\frac{1}{2}m_{{\rm eff}}^{2}\right)+\frac{i\omega}{r}\right]\varphi & \simeq 0.
\end{align}
Substituting $i\omega\rightarrow-\frac{\partial}{\partial t}$
\begin{equation}
\left[\frac{\partial}{\partial r}\frac{\partial}{\partial t}+\frac{\partial^{2}}{\partial t^{2}}+\frac{1}{2}m_{{\rm eff}}^{2}+\frac{1}{r}\frac{\partial}{\partial t}\right]\varphi=0,
\end{equation}
Therefore, $\pi=\dot{\varphi}$ obeys at the boundary $r=r_{b}$
\begin{equation}
\dot{\pi}\left(t,r_{b}\right)=-\left.\frac{\partial\pi\left(t,r\right)}{\partial r}\right|_{r=r_{b}}-\frac{1}{2}m_{{\rm eff}}^{2}\varphi\left(t,r_{b}\right)-\frac{1}{r_{b}}\pi\left(t,r_{b}\right).
\end{equation}

\bibliography{papers}

\begin{thebibliography}{34}%
\makeatletter
\providecommand \@ifxundefined [1]{%
 \@ifx{#1\undefined}
}%
\providecommand \@ifnum [1]{%
 \ifnum #1\expandafter \@firstoftwo
 \else \expandafter \@secondoftwo
 \fi
}%
\providecommand \@ifx [1]{%
 \ifx #1\expandafter \@firstoftwo
 \else \expandafter \@secondoftwo
 \fi
}%
\providecommand \natexlab [1]{#1}%
\providecommand \enquote  [1]{``#1''}%
\providecommand \bibnamefont  [1]{#1}%
\providecommand \bibfnamefont [1]{#1}%
\providecommand \citenamefont [1]{#1}%
\providecommand \href@noop [0]{\@secondoftwo}%
\providecommand \href [0]{\begingroup \@sanitize@url \@href}%
\providecommand \@href[1]{\@@startlink{#1}\@@href}%
\providecommand \@@href[1]{\endgroup#1\@@endlink}%
\providecommand \@sanitize@url [0]{\catcode `\\12\catcode `\$12\catcode
  `\&12\catcode `\#12\catcode `\^12\catcode `\_12\catcode `\%12\relax}%
\providecommand \@@startlink[1]{}%
\providecommand \@@endlink[0]{}%
\providecommand \url  [0]{\begingroup\@sanitize@url \@url }%
\providecommand \@url [1]{\endgroup\@href {#1}{\urlprefix }}%
\providecommand \urlprefix  [0]{URL }%
\providecommand \Eprint [0]{\href }%
\providecommand \doibase [0]{http://dx.doi.org/}%
\providecommand \selectlanguage [0]{\@gobble}%
\providecommand \bibinfo  [0]{\@secondoftwo}%
\providecommand \bibfield  [0]{\@secondoftwo}%
\providecommand \translation [1]{[#1]}%
\providecommand \BibitemOpen [0]{}%
\providecommand \bibitemStop [0]{}%
\providecommand \bibitemNoStop [0]{.\EOS\space}%
\providecommand \EOS [0]{\spacefactor3000\relax}%
\providecommand \BibitemShut  [1]{\csname bibitem#1\endcsname}%
\let\auto@bib@innerbib\@empty
\bibitem [{\citenamefont {Bogolyubsky}\ and\ \citenamefont
  {Makhankov}(1976)}]{Bogolyubsky:1976yu}%
  \BibitemOpen
  \bibfield  {author} {\bibinfo {author} {\bibfnamefont {I.~L.}\ \bibnamefont
  {Bogolyubsky}}\ and\ \bibinfo {author} {\bibfnamefont {V.~G.}\ \bibnamefont
  {Makhankov}},\ }\href@noop {} {\bibfield  {journal} {\bibinfo  {journal}
  {Pisma Zh. Eksp. Teor. Fiz.}\ }\textbf {\bibinfo {volume} {24}},\ \bibinfo
  {pages} {15} (\bibinfo {year} {1976})}\BibitemShut {NoStop}%
\bibitem [{\citenamefont {Gleiser}(1994)}]{Gleiser:1993pt}%
  \BibitemOpen
  \bibfield  {author} {\bibinfo {author} {\bibfnamefont {M.}~\bibnamefont
  {Gleiser}},\ }\href {\doibase 10.1103/PhysRevD.49.2978} {\bibfield  {journal}
  {\bibinfo  {journal} {Phys. Rev.}\ }\textbf {\bibinfo {volume} {D49}},\
  \bibinfo {pages} {2978} (\bibinfo {year} {1994})},\ \Eprint
  {http://arxiv.org/abs/hep-ph/9308279} {arXiv:hep-ph/9308279 [hep-ph]}
  \BibitemShut {NoStop}%
\bibitem [{\citenamefont {Copeland}\ \emph {et~al.}(1995)\citenamefont
  {Copeland}, \citenamefont {Gleiser},\ and\ \citenamefont
  {Muller}}]{Copeland:1995fq}%
  \BibitemOpen
  \bibfield  {author} {\bibinfo {author} {\bibfnamefont {E.~J.}\ \bibnamefont
  {Copeland}}, \bibinfo {author} {\bibfnamefont {M.}~\bibnamefont {Gleiser}}, \
  and\ \bibinfo {author} {\bibfnamefont {H.~R.}\ \bibnamefont {Muller}},\
  }\href {\doibase 10.1103/PhysRevD.52.1920} {\bibfield  {journal} {\bibinfo
  {journal} {Phys. Rev.}\ }\textbf {\bibinfo {volume} {D52}},\ \bibinfo {pages}
  {1920} (\bibinfo {year} {1995})},\ \Eprint
  {http://arxiv.org/abs/hep-ph/9503217} {arXiv:hep-ph/9503217 [hep-ph]}
  \BibitemShut {NoStop}%
\bibitem [{\citenamefont {McDonald}(2002)}]{McDonald:2001iv}%
  \BibitemOpen
  \bibfield  {author} {\bibinfo {author} {\bibfnamefont {J.}~\bibnamefont
  {McDonald}},\ }\href {\doibase 10.1103/PhysRevD.66.043525} {\bibfield
  {journal} {\bibinfo  {journal} {Phys. Rev.}\ }\textbf {\bibinfo {volume}
  {D66}},\ \bibinfo {pages} {043525} (\bibinfo {year} {2002})},\ \Eprint
  {http://arxiv.org/abs/hep-ph/0105235} {arXiv:hep-ph/0105235 [hep-ph]}
  \BibitemShut {NoStop}%
\bibitem [{\citenamefont {Amin}\ and\ \citenamefont
  {Shirokoff}(2010)}]{Amin:2010jq}%
  \BibitemOpen
  \bibfield  {author} {\bibinfo {author} {\bibfnamefont {M.~A.}\ \bibnamefont
  {Amin}}\ and\ \bibinfo {author} {\bibfnamefont {D.}~\bibnamefont
  {Shirokoff}},\ }\href {\doibase 10.1103/PhysRevD.81.085045} {\bibfield
  {journal} {\bibinfo  {journal} {Phys. Rev.}\ }\textbf {\bibinfo {volume}
  {D81}},\ \bibinfo {pages} {085045} (\bibinfo {year} {2010})},\ \Eprint
  {http://arxiv.org/abs/1002.3380} {arXiv:1002.3380 [astro-ph.CO]} \BibitemShut
  {NoStop}%
\bibitem [{\citenamefont {Amin}\ \emph {et~al.}(2012)\citenamefont {Amin},
  \citenamefont {Easther}, \citenamefont {Finkel}, \citenamefont {Flauger},\
  and\ \citenamefont {Hertzberg}}]{Amin:2011hj}%
  \BibitemOpen
  \bibfield  {author} {\bibinfo {author} {\bibfnamefont {M.~A.}\ \bibnamefont
  {Amin}}, \bibinfo {author} {\bibfnamefont {R.}~\bibnamefont {Easther}},
  \bibinfo {author} {\bibfnamefont {H.}~\bibnamefont {Finkel}}, \bibinfo
  {author} {\bibfnamefont {R.}~\bibnamefont {Flauger}}, \ and\ \bibinfo
  {author} {\bibfnamefont {M.~P.}\ \bibnamefont {Hertzberg}},\ }\href {\doibase
  10.1103/PhysRevLett.108.241302} {\bibfield  {journal} {\bibinfo  {journal}
  {Phys. Rev. Lett.}\ }\textbf {\bibinfo {volume} {108}},\ \bibinfo {pages}
  {241302} (\bibinfo {year} {2012})},\ \Eprint {http://arxiv.org/abs/1106.3335}
  {arXiv:1106.3335 [astro-ph.CO]} \BibitemShut {NoStop}%
\bibitem [{\citenamefont {Amin}(2013)}]{Amin:2013ika}%
  \BibitemOpen
  \bibfield  {author} {\bibinfo {author} {\bibfnamefont {M.~A.}\ \bibnamefont
  {Amin}},\ }\href {\doibase 10.1103/PhysRevD.87.123505} {\bibfield  {journal}
  {\bibinfo  {journal} {Phys. Rev.}\ }\textbf {\bibinfo {volume} {D87}},\
  \bibinfo {pages} {123505} (\bibinfo {year} {2013})},\ \Eprint
  {http://arxiv.org/abs/1303.1102} {arXiv:1303.1102 [astro-ph.CO]} \BibitemShut
  {NoStop}%
\bibitem [{\citenamefont {Takeda}\ and\ \citenamefont
  {Watanabe}(2014)}]{Takeda:2014qma}%
  \BibitemOpen
  \bibfield  {author} {\bibinfo {author} {\bibfnamefont {N.}~\bibnamefont
  {Takeda}}\ and\ \bibinfo {author} {\bibfnamefont {Y.}~\bibnamefont
  {Watanabe}},\ }\href {\doibase 10.1103/PhysRevD.90.023519} {\bibfield
  {journal} {\bibinfo  {journal} {Phys. Rev.}\ }\textbf {\bibinfo {volume}
  {D90}},\ \bibinfo {pages} {023519} (\bibinfo {year} {2014})},\ \Eprint
  {http://arxiv.org/abs/1405.3830} {arXiv:1405.3830 [astro-ph.CO]} \BibitemShut
  {NoStop}%
\bibitem [{\citenamefont {Lozanov}\ and\ \citenamefont
  {Amin}(2017)}]{Lozanov:2016hid}%
  \BibitemOpen
  \bibfield  {author} {\bibinfo {author} {\bibfnamefont {K.~D.}\ \bibnamefont
  {Lozanov}}\ and\ \bibinfo {author} {\bibfnamefont {M.~A.}\ \bibnamefont
  {Amin}},\ }\href {\doibase 10.1103/PhysRevLett.119.061301} {\bibfield
  {journal} {\bibinfo  {journal} {Phys. Rev. Lett.}\ }\textbf {\bibinfo
  {volume} {119}},\ \bibinfo {pages} {061301} (\bibinfo {year} {2017})},\
  \Eprint {http://arxiv.org/abs/1608.01213} {arXiv:1608.01213 [astro-ph.CO]}
  \BibitemShut {NoStop}%
\bibitem [{\citenamefont {Hasegawa}\ and\ \citenamefont
  {Hong}(2018)}]{Hasegawa:2017iay}%
  \BibitemOpen
  \bibfield  {author} {\bibinfo {author} {\bibfnamefont {F.}~\bibnamefont
  {Hasegawa}}\ and\ \bibinfo {author} {\bibfnamefont {J.-P.}\ \bibnamefont
  {Hong}},\ }\href {\doibase 10.1103/PhysRevD.97.083514} {\bibfield  {journal}
  {\bibinfo  {journal} {Phys. Rev.}\ }\textbf {\bibinfo {volume} {D97}},\
  \bibinfo {pages} {083514} (\bibinfo {year} {2018})},\ \Eprint
  {http://arxiv.org/abs/1710.07487} {arXiv:1710.07487 [astro-ph.CO]}
  \BibitemShut {NoStop}%
\bibitem [{\citenamefont {Antusch}\ \emph {et~al.}(2018)\citenamefont
  {Antusch}, \citenamefont {Cefala}, \citenamefont {Krippendorf}, \citenamefont
  {Muia}, \citenamefont {Orani},\ and\ \citenamefont
  {Quevedo}}]{Antusch:2017flz}%
  \BibitemOpen
  \bibfield  {author} {\bibinfo {author} {\bibfnamefont {S.}~\bibnamefont
  {Antusch}}, \bibinfo {author} {\bibfnamefont {F.}~\bibnamefont {Cefala}},
  \bibinfo {author} {\bibfnamefont {S.}~\bibnamefont {Krippendorf}}, \bibinfo
  {author} {\bibfnamefont {F.}~\bibnamefont {Muia}}, \bibinfo {author}
  {\bibfnamefont {S.}~\bibnamefont {Orani}}, \ and\ \bibinfo {author}
  {\bibfnamefont {F.}~\bibnamefont {Quevedo}},\ }\href {\doibase
  10.1007/JHEP01(2018)083} {\bibfield  {journal} {\bibinfo  {journal} {JHEP}\
  }\textbf {\bibinfo {volume} {01}},\ \bibinfo {pages} {083} (\bibinfo {year}
  {2018})},\ \Eprint {http://arxiv.org/abs/1708.08922} {arXiv:1708.08922
  [hep-th]} \BibitemShut {NoStop}%
\bibitem [{\citenamefont {Hong}\ \emph {et~al.}(2018)\citenamefont {Hong},
  \citenamefont {Kawasaki},\ and\ \citenamefont {Yamazaki}}]{Hong:2017ooe}%
  \BibitemOpen
  \bibfield  {author} {\bibinfo {author} {\bibfnamefont {J.-P.}\ \bibnamefont
  {Hong}}, \bibinfo {author} {\bibfnamefont {M.}~\bibnamefont {Kawasaki}}, \
  and\ \bibinfo {author} {\bibfnamefont {M.}~\bibnamefont {Yamazaki}},\ }\href
  {\doibase 10.1103/PhysRevD.98.043531} {\bibfield  {journal} {\bibinfo
  {journal} {Phys. Rev.}\ }\textbf {\bibinfo {volume} {D98}},\ \bibinfo {pages}
  {043531} (\bibinfo {year} {2018})},\ \Eprint
  {http://arxiv.org/abs/1711.10496} {arXiv:1711.10496 [astro-ph.CO]}
  \BibitemShut {NoStop}%
\bibitem [{\citenamefont {Kawasaki}\ \emph {et~al.}(2015)\citenamefont
  {Kawasaki}, \citenamefont {Takahashi},\ and\ \citenamefont
  {Takeda}}]{Kawasaki:2015vga}%
  \BibitemOpen
  \bibfield  {author} {\bibinfo {author} {\bibfnamefont {M.}~\bibnamefont
  {Kawasaki}}, \bibinfo {author} {\bibfnamefont {F.}~\bibnamefont {Takahashi}},
  \ and\ \bibinfo {author} {\bibfnamefont {N.}~\bibnamefont {Takeda}},\ }\href
  {\doibase 10.1103/PhysRevD.92.105024} {\bibfield  {journal} {\bibinfo
  {journal} {Phys. Rev.}\ }\textbf {\bibinfo {volume} {D92}},\ \bibinfo {pages}
  {105024} (\bibinfo {year} {2015})},\ \Eprint
  {http://arxiv.org/abs/1508.01028} {arXiv:1508.01028 [hep-th]} \BibitemShut
  {NoStop}%
\bibitem [{\citenamefont {Fodor}\ \emph {et~al.}(2006)\citenamefont {Fodor},
  \citenamefont {Forgacs}, \citenamefont {Grandclement},\ and\ \citenamefont
  {Racz}}]{Fodor:2006zs}%
  \BibitemOpen
  \bibfield  {author} {\bibinfo {author} {\bibfnamefont {G.}~\bibnamefont
  {Fodor}}, \bibinfo {author} {\bibfnamefont {P.}~\bibnamefont {Forgacs}},
  \bibinfo {author} {\bibfnamefont {P.}~\bibnamefont {Grandclement}}, \ and\
  \bibinfo {author} {\bibfnamefont {I.}~\bibnamefont {Racz}},\ }\href {\doibase
  10.1103/PhysRevD.74.124003} {\bibfield  {journal} {\bibinfo  {journal} {Phys.
  Rev.}\ }\textbf {\bibinfo {volume} {D74}},\ \bibinfo {pages} {124003}
  (\bibinfo {year} {2006})},\ \Eprint {http://arxiv.org/abs/hep-th/0609023}
  {arXiv:hep-th/0609023 [hep-th]} \BibitemShut {NoStop}%
\bibitem [{\citenamefont {Fodor}\ \emph
  {et~al.}(2009{\natexlab{a}})\citenamefont {Fodor}, \citenamefont {Forgacs},
  \citenamefont {Horvath},\ and\ \citenamefont {Mezei}}]{Fodor:2008du}%
  \BibitemOpen
  \bibfield  {author} {\bibinfo {author} {\bibfnamefont {G.}~\bibnamefont
  {Fodor}}, \bibinfo {author} {\bibfnamefont {P.}~\bibnamefont {Forgacs}},
  \bibinfo {author} {\bibfnamefont {Z.}~\bibnamefont {Horvath}}, \ and\
  \bibinfo {author} {\bibfnamefont {M.}~\bibnamefont {Mezei}},\ }\href
  {\doibase 10.1103/PhysRevD.79.065002} {\bibfield  {journal} {\bibinfo
  {journal} {Phys. Rev.}\ }\textbf {\bibinfo {volume} {D79}},\ \bibinfo {pages}
  {065002} (\bibinfo {year} {2009}{\natexlab{a}})},\ \Eprint
  {http://arxiv.org/abs/0812.1919} {arXiv:0812.1919 [hep-th]} \BibitemShut
  {NoStop}%
\bibitem [{\citenamefont {Gleiser}\ and\ \citenamefont
  {Sicilia}(2008)}]{Gleiser:2008ty}%
  \BibitemOpen
  \bibfield  {author} {\bibinfo {author} {\bibfnamefont {M.}~\bibnamefont
  {Gleiser}}\ and\ \bibinfo {author} {\bibfnamefont {D.}~\bibnamefont
  {Sicilia}},\ }\href {\doibase 10.1103/PhysRevLett.101.011602} {\bibfield
  {journal} {\bibinfo  {journal} {Phys. Rev. Lett.}\ }\textbf {\bibinfo
  {volume} {101}},\ \bibinfo {pages} {011602} (\bibinfo {year} {2008})},\
  \Eprint {http://arxiv.org/abs/0804.0791} {arXiv:0804.0791 [hep-th]}
  \BibitemShut {NoStop}%
\bibitem [{\citenamefont {Fodor}\ \emph
  {et~al.}(2009{\natexlab{b}})\citenamefont {Fodor}, \citenamefont {Forgacs},
  \citenamefont {Horvath},\ and\ \citenamefont {Mezei}}]{Fodor:2009kf}%
  \BibitemOpen
  \bibfield  {author} {\bibinfo {author} {\bibfnamefont {G.}~\bibnamefont
  {Fodor}}, \bibinfo {author} {\bibfnamefont {P.}~\bibnamefont {Forgacs}},
  \bibinfo {author} {\bibfnamefont {Z.}~\bibnamefont {Horvath}}, \ and\
  \bibinfo {author} {\bibfnamefont {M.}~\bibnamefont {Mezei}},\ }\href
  {\doibase 10.1016/j.physletb.2009.03.054} {\bibfield  {journal} {\bibinfo
  {journal} {Phys. Lett.}\ }\textbf {\bibinfo {volume} {B674}},\ \bibinfo
  {pages} {319} (\bibinfo {year} {2009}{\natexlab{b}})},\ \Eprint
  {http://arxiv.org/abs/0903.0953} {arXiv:0903.0953 [hep-th]} \BibitemShut
  {NoStop}%
\bibitem [{\citenamefont {Gleiser}\ and\ \citenamefont
  {Sicilia}(2009)}]{Gleiser:2009ys}%
  \BibitemOpen
  \bibfield  {author} {\bibinfo {author} {\bibfnamefont {M.}~\bibnamefont
  {Gleiser}}\ and\ \bibinfo {author} {\bibfnamefont {D.}~\bibnamefont
  {Sicilia}},\ }\href {\doibase 10.1103/PhysRevD.80.125037} {\bibfield
  {journal} {\bibinfo  {journal} {Phys. Rev.}\ }\textbf {\bibinfo {volume}
  {D80}},\ \bibinfo {pages} {125037} (\bibinfo {year} {2009})},\ \Eprint
  {http://arxiv.org/abs/0910.5922} {arXiv:0910.5922 [hep-th]} \BibitemShut
  {NoStop}%
\bibitem [{\citenamefont {Hertzberg}(2010)}]{Hertzberg:2010yz}%
  \BibitemOpen
  \bibfield  {author} {\bibinfo {author} {\bibfnamefont {M.~P.}\ \bibnamefont
  {Hertzberg}},\ }\href {\doibase 10.1103/PhysRevD.82.045022} {\bibfield
  {journal} {\bibinfo  {journal} {Phys. Rev.}\ }\textbf {\bibinfo {volume}
  {D82}},\ \bibinfo {pages} {045022} (\bibinfo {year} {2010})},\ \Eprint
  {http://arxiv.org/abs/1003.3459} {arXiv:1003.3459 [hep-th]} \BibitemShut
  {NoStop}%
\bibitem [{\citenamefont {Salmi}\ and\ \citenamefont
  {Hindmarsh}(2012)}]{Salmi:2012ta}%
  \BibitemOpen
  \bibfield  {author} {\bibinfo {author} {\bibfnamefont {P.}~\bibnamefont
  {Salmi}}\ and\ \bibinfo {author} {\bibfnamefont {M.}~\bibnamefont
  {Hindmarsh}},\ }\href {\doibase 10.1103/PhysRevD.85.085033} {\bibfield
  {journal} {\bibinfo  {journal} {Phys. Rev.}\ }\textbf {\bibinfo {volume}
  {D85}},\ \bibinfo {pages} {085033} (\bibinfo {year} {2012})},\ \Eprint
  {http://arxiv.org/abs/1201.1934} {arXiv:1201.1934 [hep-th]} \BibitemShut
  {NoStop}%
\bibitem [{\citenamefont {Saffin}\ \emph {et~al.}(2014)\citenamefont {Saffin},
  \citenamefont {Tognarelli},\ and\ \citenamefont {Tranberg}}]{Saffin:2014yka}%
  \BibitemOpen
  \bibfield  {author} {\bibinfo {author} {\bibfnamefont {P.~M.}\ \bibnamefont
  {Saffin}}, \bibinfo {author} {\bibfnamefont {P.}~\bibnamefont {Tognarelli}},
  \ and\ \bibinfo {author} {\bibfnamefont {A.}~\bibnamefont {Tranberg}},\
  }\href {\doibase 10.1007/JHEP08(2014)125} {\bibfield  {journal} {\bibinfo
  {journal} {JHEP}\ }\textbf {\bibinfo {volume} {08}},\ \bibinfo {pages} {125}
  (\bibinfo {year} {2014})},\ \Eprint {http://arxiv.org/abs/1401.6168}
  {arXiv:1401.6168 [hep-ph]} \BibitemShut {NoStop}%
\bibitem [{\citenamefont {Kawasaki}\ and\ \citenamefont
  {Yamada}(2014)}]{Kawasaki:2013awa}%
  \BibitemOpen
  \bibfield  {author} {\bibinfo {author} {\bibfnamefont {M.}~\bibnamefont
  {Kawasaki}}\ and\ \bibinfo {author} {\bibfnamefont {M.}~\bibnamefont
  {Yamada}},\ }\href {\doibase 10.1088/1475-7516/2014/02/001} {\bibfield
  {journal} {\bibinfo  {journal} {JCAP}\ }\textbf {\bibinfo {volume} {1402}},\
  \bibinfo {pages} {001} (\bibinfo {year} {2014})},\ \Eprint
  {http://arxiv.org/abs/1311.0985} {arXiv:1311.0985 [hep-ph]} \BibitemShut
  {NoStop}%
\bibitem [{\citenamefont {Mukaida}\ \emph {et~al.}(2017)\citenamefont
  {Mukaida}, \citenamefont {Takimoto},\ and\ \citenamefont
  {Yamada}}]{Mukaida:2016hwd}%
  \BibitemOpen
  \bibfield  {author} {\bibinfo {author} {\bibfnamefont {K.}~\bibnamefont
  {Mukaida}}, \bibinfo {author} {\bibfnamefont {M.}~\bibnamefont {Takimoto}}, \
  and\ \bibinfo {author} {\bibfnamefont {M.}~\bibnamefont {Yamada}},\ }\href
  {\doibase 10.1007/JHEP03(2017)122} {\bibfield  {journal} {\bibinfo  {journal}
  {JHEP}\ }\textbf {\bibinfo {volume} {03}},\ \bibinfo {pages} {122} (\bibinfo
  {year} {2017})},\ \Eprint {http://arxiv.org/abs/1612.07750} {arXiv:1612.07750
  [hep-ph]} \BibitemShut {NoStop}%
\bibitem [{\citenamefont {Eby}\ \emph {et~al.}(2019{\natexlab{a}})\citenamefont
  {Eby}, \citenamefont {Mukaida}, \citenamefont {Takimoto}, \citenamefont
  {Wijewardhana},\ and\ \citenamefont {Yamada}}]{Eby:2018ufi}%
  \BibitemOpen
  \bibfield  {author} {\bibinfo {author} {\bibfnamefont {J.}~\bibnamefont
  {Eby}}, \bibinfo {author} {\bibfnamefont {K.}~\bibnamefont {Mukaida}},
  \bibinfo {author} {\bibfnamefont {M.}~\bibnamefont {Takimoto}}, \bibinfo
  {author} {\bibfnamefont {L.~C.~R.}\ \bibnamefont {Wijewardhana}}, \ and\
  \bibinfo {author} {\bibfnamefont {M.}~\bibnamefont {Yamada}},\ }\href
  {\doibase 10.1103/PhysRevD.99.123503} {\bibfield  {journal} {\bibinfo
  {journal} {Phys. Rev.}\ }\textbf {\bibinfo {volume} {D99}},\ \bibinfo {pages}
  {123503} (\bibinfo {year} {2019}{\natexlab{a}})},\ \Eprint
  {http://arxiv.org/abs/1807.09795} {arXiv:1807.09795 [hep-ph]} \BibitemShut
  {NoStop}%
\bibitem [{\citenamefont {Ibe}\ \emph {et~al.}(2019)\citenamefont {Ibe},
  \citenamefont {Kawasaki}, \citenamefont {Nakano},\ and\ \citenamefont
  {Sonomoto}}]{Ibe:2019vyo}%
  \BibitemOpen
  \bibfield  {author} {\bibinfo {author} {\bibfnamefont {M.}~\bibnamefont
  {Ibe}}, \bibinfo {author} {\bibfnamefont {M.}~\bibnamefont {Kawasaki}},
  \bibinfo {author} {\bibfnamefont {W.}~\bibnamefont {Nakano}}, \ and\ \bibinfo
  {author} {\bibfnamefont {E.}~\bibnamefont {Sonomoto}},\ }\href {\doibase
  10.1007/JHEP04(2019)030} {\bibfield  {journal} {\bibinfo  {journal} {JHEP}\
  }\textbf {\bibinfo {volume} {04}},\ \bibinfo {pages} {030} (\bibinfo {year}
  {2019})},\ \Eprint {http://arxiv.org/abs/1901.06130} {arXiv:1901.06130
  [hep-ph]} \BibitemShut {NoStop}%
\bibitem [{\citenamefont {Oll{\'e}}\ \emph {et~al.}(2019)\citenamefont
  {Oll{\'e}}, \citenamefont {Pujol{\`a}s},\ and\ \citenamefont
  {Rompineve}}]{Olle:2019kbo}%
  \BibitemOpen
  \bibfield  {author} {\bibinfo {author} {\bibfnamefont {J.}~\bibnamefont
  {Oll{\'e}}}, \bibinfo {author} {\bibfnamefont {O.}~\bibnamefont
  {Pujol{\`a}s}}, \ and\ \bibinfo {author} {\bibfnamefont {F.}~\bibnamefont
  {Rompineve}},\ }\href@noop {} {\  (\bibinfo {year} {2019})},\ \Eprint
  {http://arxiv.org/abs/1906.06352} {arXiv:1906.06352 [hep-ph]} \BibitemShut
  {NoStop}%
\bibitem [{\citenamefont {Antusch}\ \emph {et~al.}(2019)\citenamefont
  {Antusch}, \citenamefont {Cefal^^c3^^a0},\ and\ \citenamefont
  {Torrent^^c3^^ad}}]{Antusch:2019qrr}%
  \BibitemOpen
  \bibfield  {author} {\bibinfo {author} {\bibfnamefont {S.}~\bibnamefont
  {Antusch}}, \bibinfo {author} {\bibfnamefont {F.}~\bibnamefont
  {Cefal^^c3^^a0}}, \ and\ \bibinfo {author} {\bibfnamefont {F.}~\bibnamefont
  {Torrent^^c3^^ad}},\ }\href@noop {} {\  (\bibinfo {year} {2019})},\ \Eprint
  {http://arxiv.org/abs/1907.00611} {arXiv:1907.00611 [hep-ph]} \BibitemShut
  {NoStop}%
\bibitem [{\citenamefont {Zhou}\ \emph {et~al.}(2013)\citenamefont {Zhou},
  \citenamefont {Copeland}, \citenamefont {Easther}, \citenamefont {Finkel},
  \citenamefont {Mou},\ and\ \citenamefont {Saffin}}]{Zhou:2013tsa}%
  \BibitemOpen
  \bibfield  {author} {\bibinfo {author} {\bibfnamefont {S.-Y.}\ \bibnamefont
  {Zhou}}, \bibinfo {author} {\bibfnamefont {E.~J.}\ \bibnamefont {Copeland}},
  \bibinfo {author} {\bibfnamefont {R.}~\bibnamefont {Easther}}, \bibinfo
  {author} {\bibfnamefont {H.}~\bibnamefont {Finkel}}, \bibinfo {author}
  {\bibfnamefont {Z.-G.}\ \bibnamefont {Mou}}, \ and\ \bibinfo {author}
  {\bibfnamefont {P.~M.}\ \bibnamefont {Saffin}},\ }\href {\doibase
  10.1007/JHEP10(2013)026} {\bibfield  {journal} {\bibinfo  {journal} {JHEP}\
  }\textbf {\bibinfo {volume} {10}},\ \bibinfo {pages} {026} (\bibinfo {year}
  {2013})},\ \Eprint {http://arxiv.org/abs/1304.6094} {arXiv:1304.6094
  [astro-ph.CO]} \BibitemShut {NoStop}%
\bibitem [{\citenamefont {Antusch}\ \emph {et~al.}(2017)\citenamefont
  {Antusch}, \citenamefont {Cefala},\ and\ \citenamefont
  {Orani}}]{Antusch:2016con}%
  \BibitemOpen
  \bibfield  {author} {\bibinfo {author} {\bibfnamefont {S.}~\bibnamefont
  {Antusch}}, \bibinfo {author} {\bibfnamefont {F.}~\bibnamefont {Cefala}}, \
  and\ \bibinfo {author} {\bibfnamefont {S.}~\bibnamefont {Orani}},\ }\href
  {\doibase 10.1103/PhysRevLett.120.219901, 10.1103/PhysRevLett.118.011303}
  {\bibfield  {journal} {\bibinfo  {journal} {Phys. Rev. Lett.}\ }\textbf
  {\bibinfo {volume} {118}},\ \bibinfo {pages} {011303} (\bibinfo {year}
  {2017})},\ \bibinfo {note} {[Erratum: Phys. Rev.
  Lett.120,no.21,219901(2018)]},\ \Eprint {http://arxiv.org/abs/1607.01314}
  {arXiv:1607.01314 [astro-ph.CO]} \BibitemShut {NoStop}%
\bibitem [{\citenamefont {Lozanov}\ and\ \citenamefont
  {Amin}(2019)}]{Lozanov:2019ylm}%
  \BibitemOpen
  \bibfield  {author} {\bibinfo {author} {\bibfnamefont {K.~D.}\ \bibnamefont
  {Lozanov}}\ and\ \bibinfo {author} {\bibfnamefont {M.~A.}\ \bibnamefont
  {Amin}},\ }\href {\doibase 10.1103/PhysRevD.99.123504} {\bibfield  {journal}
  {\bibinfo  {journal} {Phys. Rev.}\ }\textbf {\bibinfo {volume} {D99}},\
  \bibinfo {pages} {123504} (\bibinfo {year} {2019})},\ \Eprint
  {http://arxiv.org/abs/1902.06736} {arXiv:1902.06736 [astro-ph.CO]}
  \BibitemShut {NoStop}%
\bibitem [{\citenamefont {Eby}\ \emph {et~al.}(2019{\natexlab{b}})\citenamefont
  {Eby}, \citenamefont {Leembruggen}, \citenamefont {Street}, \citenamefont
  {Suranyi},\ and\ \citenamefont {Wijewardhana}}]{Eby:2019ntd}%
  \BibitemOpen
  \bibfield  {author} {\bibinfo {author} {\bibfnamefont {J.}~\bibnamefont
  {Eby}}, \bibinfo {author} {\bibfnamefont {M.}~\bibnamefont {Leembruggen}},
  \bibinfo {author} {\bibfnamefont {L.}~\bibnamefont {Street}}, \bibinfo
  {author} {\bibfnamefont {P.}~\bibnamefont {Suranyi}}, \ and\ \bibinfo
  {author} {\bibfnamefont {L.~C.~R.}\ \bibnamefont {Wijewardhana}},\
  }\href@noop {} {\  (\bibinfo {year} {2019}{\natexlab{b}})},\ \Eprint
  {http://arxiv.org/abs/1905.00981} {arXiv:1905.00981 [hep-ph]} \BibitemShut
  {NoStop}%
\bibitem [{\citenamefont {Kasuya}\ \emph {et~al.}(2003)\citenamefont {Kasuya},
  \citenamefont {Kawasaki},\ and\ \citenamefont {Takahashi}}]{Kasuya:2002zs}%
  \BibitemOpen
  \bibfield  {author} {\bibinfo {author} {\bibfnamefont {S.}~\bibnamefont
  {Kasuya}}, \bibinfo {author} {\bibfnamefont {M.}~\bibnamefont {Kawasaki}}, \
  and\ \bibinfo {author} {\bibfnamefont {F.}~\bibnamefont {Takahashi}},\ }\href
  {\doibase 10.1016/S0370-2693(03)00344-7} {\bibfield  {journal} {\bibinfo
  {journal} {Phys. Lett.}\ }\textbf {\bibinfo {volume} {B559}},\ \bibinfo
  {pages} {99} (\bibinfo {year} {2003})},\ \Eprint
  {http://arxiv.org/abs/hep-ph/0209358} {arXiv:hep-ph/0209358 [hep-ph]}
  \BibitemShut {NoStop}%
\bibitem [{\citenamefont {Mukaida}\ and\ \citenamefont
  {Takimoto}(2014)}]{Mukaida:2014oza}%
  \BibitemOpen
  \bibfield  {author} {\bibinfo {author} {\bibfnamefont {K.}~\bibnamefont
  {Mukaida}}\ and\ \bibinfo {author} {\bibfnamefont {M.}~\bibnamefont
  {Takimoto}},\ }\href {\doibase 10.1088/1475-7516/2014/08/051} {\bibfield
  {journal} {\bibinfo  {journal} {JCAP}\ }\textbf {\bibinfo {volume} {1408}},\
  \bibinfo {pages} {051} (\bibinfo {year} {2014})},\ \Eprint
  {http://arxiv.org/abs/1405.3233} {arXiv:1405.3233 [hep-ph]} \BibitemShut
  {NoStop}%
\bibitem [{\citenamefont {Engquist}\ and\ \citenamefont
  {Majda}(1977)}]{d4443ca6ddba410f8a155c4dd67936d0}%
  \BibitemOpen
  \bibfield  {author} {\bibinfo {author} {\bibfnamefont {B.}~\bibnamefont
  {Engquist}}\ and\ \bibinfo {author} {\bibfnamefont {A.}~\bibnamefont
  {Majda}},\ }\href {\doibase 10.1090/S0025-5718-1977-0436612-4} {\bibfield
  {journal} {\bibinfo  {journal} {Mathematics of Computation}\ }\textbf
  {\bibinfo {volume} {31}},\ \bibinfo {pages} {629} (\bibinfo {year}
  {1977})}\BibitemShut {NoStop}%
\end{thebibliography}%


%
\end{document}